\documentclass[
    prl, twocolumn, superscriptaddress, nofootinbib,nobibnotes, amsmath, amssymb,
    aps, floatfix, preprintnumbers, 10pt
]{revtex4-2}
\usepackage{tabularx}
\usepackage{setspace}
\usepackage[utf8]{inputenc}
\usepackage[dvips]{graphicx}
\usepackage[dvipsnames]{xcolor}
\usepackage{float}
\usepackage[english]{babel}
\babeladjust{autoload.bcp47 = on}
\definecolor{CiteBlue}{RGB}{45,52,151}
\usepackage[
    colorlinks=true,
    linkcolor=CiteBlue,
    urlcolor=CiteBlue,
    citecolor=CiteBlue
]{hyperref}
\usepackage{aas_macros}
\usepackage[capitalise]{cleveref}
\usepackage{times}
\usepackage{bm}

\newcommand{\fref}[1]{Fig.~\ref{#1}}
\newcommand{\tref}[1]{Table~\ref{#1}}
\newcommand{\eref}[1]{Eq.~(\ref{#1})}
\newcommand{\erefs}[2]{Eqs.~(\ref{#1})~and~(\ref{#2})}
\newcommand{\rref}[1]{Ref.~\cite{#1}}
\newcommand{\SQWARE}[1]{\text{SQWARE}}
\newcommand{\sref}[1]{Sec.~\ref{#1}}

\newcommand{\gagg}{g_{\mathrm{a}\gamma\gamma}}
\newcommand{\Evec}{\mathbf{E}}
\newcommand{\Bvec}{\mathbf{B}}
\newcommand{\Dvec}{\mathbf{D}}
\newcommand{\Hvec}{\mathbf{H}}
\newcommand{\Jvec}{\mathbf{J}}
\newcommand{\rvec}{\mathbf{r}}
\newcommand{\kvec}{\mathbf{k}}
\newcommand{\nvec}{\mathbf{n}}
\newcommand{\zvec}{\mathbf{0}}
\newcommand{\evec}{\mathbf{e}}
\newcommand{\dvec}{{\bm \nabla}}
\newcommand{\pprime}{{\prime\prime}}

\usepackage[normalem]{ulem}

\usepackage{verbatim}
\usepackage{booktabs}
\usepackage{boldline}
\usepackage{threeparttable}
\usepackage{nicematrix}
\usepackage{multirow}
\usepackage{amsmath}
\usepackage[en-GB]{datetime2}
\usepackage{upgreek}
\usepackage{lineno}

\begin{document}





\title{Quantum Semiconductor Heterostructures for meV Axion Dark Matter Detection}

\author{Jaanita~Mehrani}
\email{jaanita.s.mehrani@rice.edu}
\affiliation{Applied Physics Graduate Program, Smalley-Curl Institute, Rice University, Houston, Texas 77005, USA}
\affiliation{Department of Electrical and Computer Engineering, Rice University, Houston, Texas 77005, USA}

\author{Tao~Xu}
\email{tao.xu@ou.edu}
\affiliation{Homer L. Dodge Department of Physics and Astronomy, University of Oklahoma, Norman, Oklahoma 73019, USA}

\author{Andrey~Baydin}
\email{andrey.baydin@rice.edu}
\affiliation{Department of Electrical and Computer Engineering, Rice University, Houston, Texas 77005, USA}
\affiliation{Smalley-Curl Institute, Rice University, Houston, Texas 77005, USA}

\author{Michael~J.~Manfra}
\email{mmanfra@purdue.edu}
\affiliation{Department of Physics and Astronomy, Purdue University, West Lafayette, Indiana 47907, USA}
\affiliation{Birck Nanotechnology Center, Purdue University, West Lafayette, Indiana 47907, USA}
\affiliation{School of Electrical and Computer Engineering, Purdue University, West Lafayette, Indiana 47907, USA}
\affiliation{School of Materials Engineering, Purdue University, West Lafayette, Indiana 47907, USA}

\author{Henry~O.~Everitt}
\email{henry.everitt@rice.edu}
\affiliation{Department of Electrical and Computer Engineering, Rice University, Houston, Texas 77005, USA}
\affiliation{Smalley-Curl Institute, Rice University, Houston, Texas 77005, USA}
\affiliation{Department of Physics and Astronomy, Rice University, Houston, Texas 77005, USA}
\affiliation{DEVCOM Army Research Laboratory-South, 6100 Main Street, Houston, Texas, 77005, USA}

\author{Andrew~J.~Long}
\email{andrewjlong@rice.edu}
\affiliation{Department of Physics and Astronomy, Rice University, Houston, Texas 77005, USA}

\author{Kuver~Sinha}
\email{kuver.sinha@ou.edu}
\affiliation{Homer L. Dodge Department of Physics and Astronomy, University of Oklahoma, Norman, OK 73019, USA}

\author{Junichiro~Kono}
\email{kono@rice.edu}
\affiliation{Department of Electrical and Computer Engineering, Rice University, Houston, Texas 77005, USA}
\affiliation{Smalley-Curl Institute, Rice University, Houston, Texas 77005, USA}
\affiliation{Department of Physics and Astronomy, Rice University, Houston, Texas 77005, USA}
\affiliation{Department of Materials Science and NanoEngineering, Rice University, Houston, Texas 77005, USA}
\affiliation{Rice Advanced Materials Institute, Rice University, Houston, Texas 77005, USA}

\author{Shengxi~Huang}
\email{shengxi.huang@rice.edu}
\affiliation{Department of Electrical and Computer Engineering, Rice University, Houston, Texas 77005, USA}
\affiliation{Smalley-Curl Institute, Rice University, Houston, Texas 77005, USA}
\affiliation{Department of Materials Science and NanoEngineering, Rice University, Houston, Texas 77005, USA}
\affiliation{Rice Advanced Materials Institute, Rice University, Houston, Texas 77005, USA}
\affiliation{Department of Bioengineering, Rice University, Houston, Texas 77005, USA}

\begin{abstract}
We propose a novel strategy and a new class of detectors for the direct detection of axion dark matter in the meV mass range, based on resonantly enhanced axion-photon conversion through the inverse Primakoff effect in engineered radiometers composed of quantum semiconductor heterostructures. Semiconductor-Quantum-Well Axion Radiometer Experiments (\SQWARE{}s) are multiple quantum well structures forming magnetoplasmonic cavities, containing high-mobility two-dimensional electron gases, realizing tunable epsilon-near-zero resonances in the terahertz frequency range. By controlling the orientation of the cavity within a strong external magnetic field, both the resonance frequency and the axion-induced current are optimized \textit{in~situ}, enabling efficient scanning across a broad mass range without the need for complex mechanical adjustments. The axion-induced electromagnetic signal radiatively emitted from the cavity is then detected by a photodetector. We present the theoretical basis for resonant enhancement, detail the experimental design and benchmarks through extensive simulations, project the sensitivity of an example \SQWARE{} for several realistic configurations, and demonstrate the modularity and flexibility of the design to fit reasonably with any lab's existing capabilities and target unique axion mass ranges. Our results demonstrate that the \SQWARE{}s can probe the well-motivated quantum chromodynamics axion parameter space and close a critical gap in direct searches at meV masses.
\end{abstract}
\maketitle
\setcounter{equation}{0}
\textit{Introduction}---The axion, originally proposed to solve the strong charge-parity problem in quantum chromodynamics (QCD), remains a leading candidate for dark matter (DM)~\cite{preskill_cosmology_1983,abbott_cosmological_1983,dine_not-so-harmless_1983,adams_axion_2023}. Axionlike particles, which we will also call axions, appear ubiquitous in string theory~\cite{reece_tasi_2023}. Searches probing various portions of the axion mass $m_\mathrm{a}$ use photons as the primary probe through the axion-photon coupling $\gagg$ in planned colliders, beam dumps, astrophysical systems, and an array of tabletop experiments~\cite{ohare_cosmology_2024}. 
Masses in the meV range lie on the edge of favorability for postinflationary production of axions and correspond to maximal misalignment in the preinflationary scenario
~\cite{navas_review_2024, luzio_landscape_2020}.

Direct detection of axion DM at meV masses poses a significant experimental challenge. Traditional vacuum cavity haloscopes like ADMX, CAPP, and QUAX are highly sensitive at microwave frequencies, with high quality factors $Q\sim 10^4-10^5$ but narrow tuning bandwidth~\cite{semertzidis_axion_2022}. At THz frequencies corresponding to meV masses, however, the quality factor and cavity volume drop significantly~\cite{reuter_theory_1948}. Metallic cavities like Fabry-Perot have experimentally demonstrated up to $Q\sim 89$~\cite{mavrona_thz_2021}, and while Tamm and photonic crystal cavities reach $Q\sim 10^3-10^4$, they have low form factor, further reducing the signal~\cite{tu_tamm-cavity_2024, okamoto_terahertz_2017}. 
Mechanical tuning mechanisms involving dielectric rods and/or multicavity systems become difficult to implement with smaller structures due to shorter wavelengths and in smaller sample spaces that are common in high field magnets~\cite{brandt_national_2001}. Novel strategies, including dielectric haloscopes~\cite{millar_dielectric_2017, group_dielectric_2017} and plasmonic cavity designs~\cite{kowitt_tunable_2023, balafendiev_wire_2022, lawson_tunable_2019}, have extended the reach of axion searches, but mechanical tuning and structural constraints limit their scalability at higher frequencies. Meanwhile, detectors based on material resonances, such as phonon-polariton~\cite{mitridate_detectability_2020, marsh_axion_2023} and axion quasiparticle devices~\cite{li_dynamical_2010,marsh_proposal_2019,schutte-engel_axion_2021, qiu_observation_2025}, are restricted by the scarcity of well-characterized, low-loss materials with tunable resonances at meV energies~\cite{kafesaki_thz_2014}. Broadband approaches like Broadband Reflector Experiment for Axion Detection (BREAD), which uses parabolic reflectors, are promising but require a very large detector volume~\cite{liu_broadband_2022}.

Here, we propose a novel detection strategy, the Semiconductor-Quantum-Well Axion Radiometer Experiments (\SQWARE{}s), that leverages advances in quantum semiconductor heterostructures to overcome these obstacles. 
Our proposed experimental design utilizes a magnetoplasmonic cavity based on high-mobility two-dimensional electron gases (2DEGs) in a multiple quantum well (MQW) structure in a strong magnetic field engineered with a tunable epsilon-near-zero (ENZ) response in the THz band, where the axion-photon conversion signal can be resonantly and electromagnetically enhanced. In contrast to previous methods, the resonant frequency is scanned \textit{in~situ} simply by reorienting the MQW in the magnetic field. This approach enables efficient exploration of the previously inaccessible meV axion mass window, with projected sensitivities that can approach the well-motivated QCD axion parameter space under the Dine-Fischler-Srednicki-Zhitnitsky and Kim-Shifman-Vainshtein-Zakharov models~\cite{luzio_landscape_2020}. The modular design also allows labs with different resource availabilities, including weaker magnets and alternative growth methods and semiconductor materials, to target unique mass ranges. The combination of using known high-quality materials with wide tunability and flexible, scalable fabrication methods marks a major step toward closing the experimental gap in axion DM searches. 
 \begin{figure}[h!]
 \includegraphics[width=0.42\textwidth]{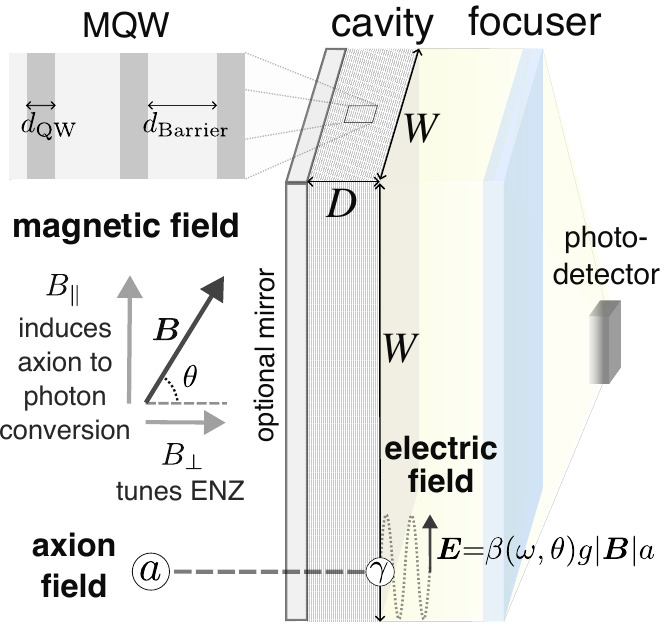}
 \caption{\label{fig:schematic}
 Schematic of a possible \SQWARE{} detector. A magnetoplasmonic cavity with area $W^2$ and thickness $D$ is an MQW structure consisting of high-mobility 2DEGs in quantum wells with thickness $d_{\text{QW}}$ and insulating dielectric barriers with thickness $d_{\text{Barrier}}$. An applied magnetic field $\boldsymbol{B}$ with tilt angle $\theta$ induces electromagnetic radiation from the axion DM (via component $B_\parallel$) and sweeps the ENZ resonance frequency (via $B_\perp$). Electromagnetic radiation from the surface of the cavity is focused onto a photodetector.} 
 \end{figure}
 
The \SQWARE{} concept consists of an MQW structure functioning as a magnetoplasmonic cavity with centimeter-scale thickness and diameter placed in a static magnetic field. The relevant physical scales (electromagnetic wavelength $\lambda \sim \mathrm{mm}$, frequency $f \sim \mathrm{THz}$, and axion mass $m_\mathrm{a} \sim \mathrm{meV}$, for example) are well-matched to the engineered ENZ resonance of the MQW, where the permittivity approaches zero. Under these conditions, the axion-induced electric field inside the material is resonantly enhanced, substantially increasing the emitted electromagnetic power on the order of $\sim10^4$, which can be measured with a photodetector coupled to the device. Crucially, the ENZ resonance is tunable. By varying the orientation of the MQW within the external magnetic field, one can independently optimize the magnetic field components due to the anisotropic permittivity tensor of the material: the component parallel to the sample controls the axion-induced current and thus the signal strength, while the component perpendicular to the sample tunes the ENZ frequency, allowing for mass scanning. This dual control enables efficient exploration of the $(m_\mathrm{a}, \gagg)$ parameter space without complicated mechanical adjustment. Figure~\ref{fig:schematic} illustrates the core \SQWARE{} concept. The remainder of this Letter details the underlying physical principles and presents projected sensitivities to axion DM across the meV mass window for an example ideal \SQWARE{} configuration.

\textit{General Concept}---The axion-photon coupling modifies Maxwell's equations by introducing a source term proportional to the axion field and external magnetic field; see Sec.~I~A in Supplemental Material (SM)~\cite{supplement}. For a nonrelativistic axion DM background, the axion field can be approximated as spatially homogeneous on the scale of the experiment, $a(t) = a_0 e^{-im_\mathrm{a} t}$, with amplitude $a_0 = \sqrt{2\rho_\mathrm{a}}/m_\mathrm{a}$, where $\rho_\mathrm{a}$ is the local axion DM density. In a medium with a complex electric permittivity tensor $\varepsilon(\omega)$ (real part $\varepsilon'$ and imaginary part $\varepsilon''$), the induced electric field with vector amplitude $\boldsymbol{E}$ oscillates at the axion Compton frequency $\omega_\mathrm{a} = m_\mathrm{a}$, 
\begin{equation}
\varepsilon(\omega) \boldsymbol{E} {e^{-i\omega_\mathrm{a} t}} = -\gagg  \boldsymbol{B} a_0 {e^{-im_\mathrm{a} t}},
\label{eq:axion-maxwell}
\end{equation}
where $\boldsymbol{B}$ is the applied dc magnetic field vector.

The axion-induced electric field can be dramatically amplified in media where the permittivity approaches zero. In this regime, the field in the medium $E_{{\mathrm{med}}}$ is enhanced by a  
factor,
\begin{equation}
    \frac{E_{\mathrm{med}}}{E_{\mathrm{vac}}} = \frac{1}{|\varepsilon|},
\end{equation}
where $E_{\mathrm{vac}} = \gagg|\boldsymbol{B}| a_0$ is the axion-induced field in vacuum~\cite{millar_dielectric_2017}. Thus, minimizing $|\varepsilon|$ produces a large electromagnetic response to the axion; see Sec.~I~B in SM~\cite{supplement}. This enhancement is central to the \SQWARE{} concept: by engineering a tunable ENZ resonance in a semiconductor magnetoplasma, the signal can be resonantly boosted and efficiently measured by a photodetector. \\

\textit{Boost Factor}---Whereas the homogeneous electric field in an infinite medium is enhanced by a factor of $1/|\varepsilon|$, the boost factor is somewhat reduced when considering the propagating radiation emitted from a finite volume in a real experiment.
The \SQWARE{} design uses an anisotropic magnetoplasmonic cavity at an angle $\theta$ relative to the applied magnetic field to generate a tunable ENZ mode in one of the two chiral permittivity components.
Following similar steps as~\rref{millar_dielectric_2017}, the boost factor $\beta$ in the radiated field from such a cavity is 
\begin{equation}
\beta(\omega, \theta) = \bigg|-\frac{ (1-\varepsilon)\sin(\Delta/2)  }
{\varepsilon\sin(\Delta/2) + i\sqrt{\varepsilon} \cos(\Delta/2)  }\bigg| \frac{\sin\theta}{\sqrt{2}} ,
\label{boostonedisk}
\end{equation}
where $\Delta = 2\pi D/\lambda_\mathrm{med}$ is the phase depth of a cavity with thickness $D$, $\lambda_\mathrm{med}= \lambda_\mathrm{vac}/n_\mathrm{med}$ is the in-medium wavelength, $n_\mathrm{med}= \sqrt{\varepsilon}$ is the complex refractive index, and $\lambda_\mathrm{vac} = 2\pi c/\omega$ is the vacuum wavelength. 

\begin{figure}[h!]
    \centering    \includegraphics[width=0.95\linewidth]{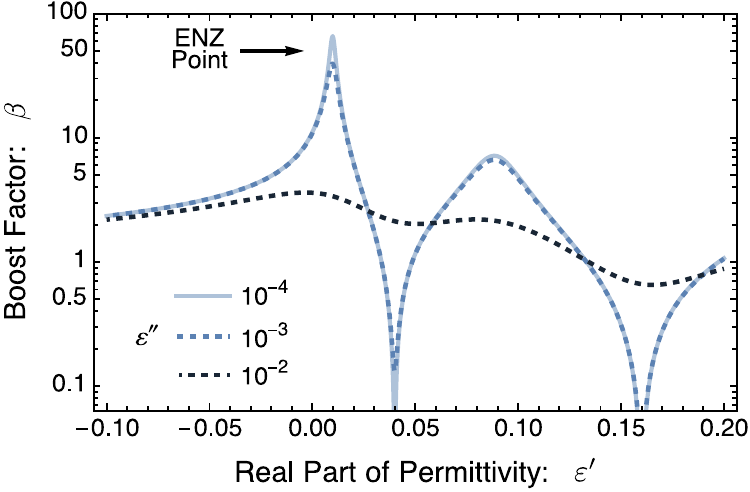}
    \caption{\label{Single cavity function of frequency}
    The boost factor $\beta$, given by~\eref{boostonedisk}, from a plasmonic cavity with a fixed thickness $D= 5 \lambda_\mathrm{vac}$ as a function of the complex permittivity $\varepsilon = \varepsilon^\prime + i \varepsilon^{\prime\prime}$, evaluated at $\theta=90^{\circ}$. The resonant boost factor occurs at the ENZ point, and decreasing loss (or equivalently $\varepsilon''$) maximizes the achievable boost, which can be quite large.}
\end{figure}

The boost factor is now maximized at a finite real part of the permittivity $\varepsilon'$ (still ENZ). For negligible $\mathrm{Im}[\Delta]$, resonant enhancement occurs when $\mathrm{Re}[\Delta]\approx M\pi$, where $M$ is an odd integer, and the boost factor on resonance is $(1-\varepsilon)\sin\theta/\sqrt{2}\varepsilon$. The boost factor is modified with the addition of a mirror to redirect the radiation from both sides of the magnetoplasmonic cavity toward the photodetector. See Sec.~I~D and I~E in SM~\cite{supplement} for the derivation of the boost factors~\cite{supplement}. A critical parameter governing the enhancement is the loss, which is proportional to the imaginary part of the permittivity, $\varepsilon''$. High loss broadens and suppresses the resonance, significantly reducing $\beta$ and the detectable signal, illustrated in ~\fref{Single cavity function of frequency}. \SQWARE{} must be designed with optimized material parameters to achieve low-loss, high-boost performance in the ENZ regime.

\textit{MQW}---The \SQWARE{} detector concept exploits well-characterized engineered MQWs, such as a GaAs/AlGaAs system with ultrahigh mobility formed from molecular beam epitaxy (MBE), a well-developed and high-quality fabrication method, to create a magnetoplasmonic cavity with sharp, tunable ENZ response at the frequency of interest. Each quantum well consists of a thin layer of a doped degenerate semiconductor (thickness $d_\text{QW}$) separated by thicker insulating dielectric barrier layers (thickness $d_{\text{Barrier}}$), forming periodic 2DEGs with characteristic electron surface density $n_\mathrm{e}$ and scattering time $\tau$ that is directly proportional to the material's intrinsic mobility $\mu_\mathrm{e} = e \tau / m^*$, both determined by currently achievable values in high-quality (ultrahigh-moblity) systems. $n_\mathrm{e}$ is set by an atomically thin silicon doping layer providing electrons to the nearby quantum well, while $\tau$ is governed by impurities (which are on the order $10^{13}$\,cm$^{-3}$ in high-quality samples) and thermal phonons~\cite{manfra_molecular_2014}. Cryogenic temperatures ($T = 0.3$\,K) are necessary to ensure negligible thermal noise as well as long scattering times,  which results in lower loss and a stronger, sharper resonance. The collective electronic response of a 2DEG, which can occur on timescales faster than the axion coherence time~\cite{kono_picosecond_1999}, may be described by the Drude model, from which a frequency and magnetic field dependent electric permittivity tensor may be obtained~\cite{kelly_physics_1985, baer_two-dimensional_2015}.  
When subjected to an external magnetic field $\boldsymbol{B}$, cyclotron resonance allows for multiple polarized ENZ points in the permittivity tensor $\varepsilon$ within the relevant axion mass window.

\begin{figure*}[ht]
    \centering
    {\includegraphics[height=0.278\linewidth]{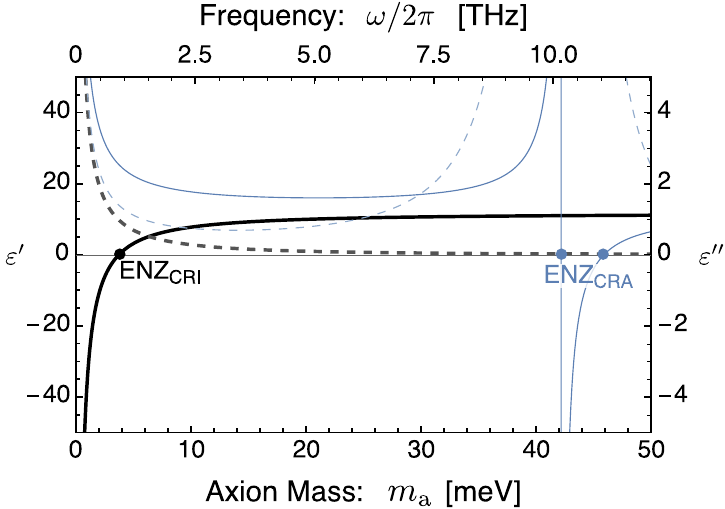}}
    \hspace{0.1cm}
    {\includegraphics[height=0.278\linewidth]{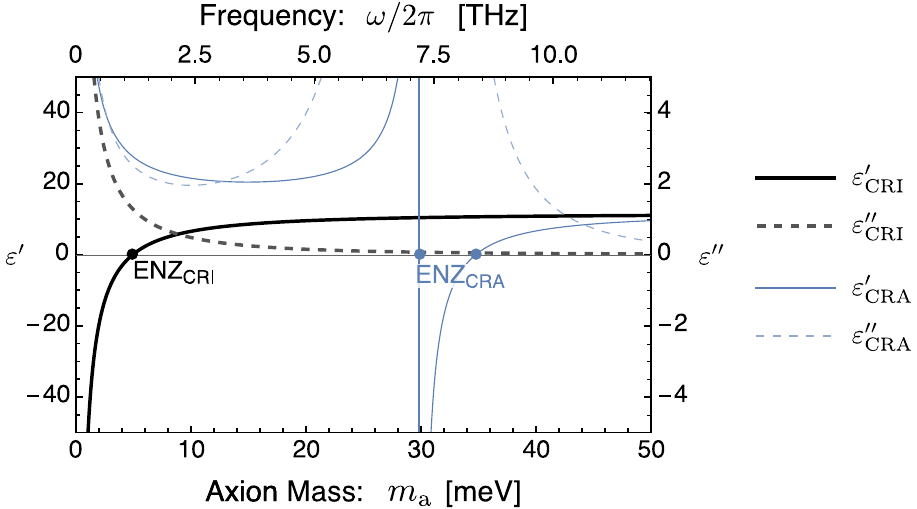}}
    \caption{{The CRI and CRA permittivity components in a single GaAs-AlGaAs quantum well as a function of electromagnetic frequency for two magnetic fields are plotted. We plot the real and imaginary parts, $\varepsilon = \varepsilon^\prime + i \varepsilon^\pprime$, evaluated at $\kvec = \zvec$ and $\omega = m_\mathrm{a}$. We take the \SQWARE{} Config.~2's material parameters, with magnetic field $|\Bvec_\mathrm{ext}| = 36~\mathrm{T}$ orientated at angle $\theta = 45^{\circ}$ (left) and $\theta = 60^{\circ}$ (right) from the 2DEG's surface normal. A larger angle and normal magnetic field $B_\perp$ shifts the CRI's ENZ point to higher frequency and vice versa for CRA.}
    }
    \label{qwcrpermm}
\end{figure*}

The key tuning parameter is the cyclotron frequency, $\omega_\mathrm{c} = eB_\perp/m^*$, where $B_\perp = |\boldsymbol{B}| \cos\theta$ is the magnetic field component normal to the quantum well surface, $e$ is the electron charge, and $m^*$ is the effective electron mass~\cite{sze_physics_2021}. 
The independent chiral components of the permittivity tensor, cyclotron resonance inactive (CRI) and cyclotron resonance active (CRA), are normal to the stacking direction and respond uniquely to $B_\perp$. 
Typically, the CRI component is most relevant for detection as it exhibits a sharper ENZ resonance with minimal loss, while the CRA mode is strongly damped by absorption represented by a large imaginary permittivity near resonance. The CRI permittivity in the quantum well is given by~\cite{hilton_cyclotron_2012, li_vacuum_2018} (see Sec.~I~C in SM~\cite{supplement})
\begin{equation}
    \varepsilon_\text{CRI} = \varepsilon_\text{bg} +  \frac{in_\mathrm{e}e^2\tau}{m^*\omega {d_\text{QW}} \left [ 1 - i (\omega + \omega_\mathrm{c}) \tau \right ] }
\end{equation}
where $\varepsilon_\text{bg}$ is the dielectric constant of the doped semiconductor---for example, GaAs~\cite{adachi_gaas_1985}. 
As the cyclotron resonance frequency $\omega_\mathrm{c}$ increases as the result of an increasing $B_\perp$, the ENZ point will shift toward lower frequencies and the signal will resonate at lower axion masses, leading to the unique scaling $B_\perp \propto 1/m_\mathrm{a}$, where $m_\mathrm{a} = \omega$ when $\omega \ll \omega_\mathrm{c}$, $ \omega  \ll \sigma_0$, and $1\ll \omega_c\tau$. See \fref{qwcrpermm} for the complex permittivity tensor and ENZ points for a single GaAs-AlGaAs quantum well as a function of frequency at two different magnetic fields.

From effective medium theory (EMT) under the Maxwell-Garnett approximation~\cite{haija_review_2011,ferrari_hyperbolic_2015,popov_surface_2018,wu_frequency-tunable_2022}, the overall permittivity components normal to the stacking direction ($\varepsilon_\text{eff, CRI}$ and $\varepsilon_\text{eff, CRA}$) are the weighted average of the quantum well and barrier layers, $\varepsilon_\text{QW}$ (either $\varepsilon_\text{CRI}$ or $\varepsilon_\text{CRA}$) and $\varepsilon_\text{Barrier}$, the isotropic dielectric constant of the barrier, e.g., AlGaAs~\cite{talghader_thermal_1995, gehrsitz_refractive_2000},
\begin{equation}
\varepsilon_\text{eff, CRI} = \frac{\varepsilon_{\text{Barrier}} d_{\text{Barrier}} + \varepsilon_{\text{CRI}} d_{\text{QW}}}{d_{\text{Barrier}} + d_{\text{QW}}}.
\label{eq:effmedperm}
\end{equation}

EMT is valid when individual layer thicknesses are small compared to in-medium electromagnetic wavelengths (see End Matter) and the doped semiconductor layers remain 2DEGs~\cite{ando_electronic_1982, gilbertson_dimensional_2009}. Equation~(\ref{eq:effmedperm}) substitutes the permittivity in \eref{boostonedisk} to calculate signal boost. To achieve effective ENZ, the quantum well's permittivity must approach a large negative value to compensate for the thicker barrier's positive permittivity. THz ENZ has been observed in MQWs~\cite{li_vacuum_2018}. All parameters in the boost $\beta(\omega, \theta)$ are fixed, except for the magnetic field orientation angle $\theta$ and the axion or photon frequency $\omega$. The signal enhancement is maximized at optimal configurations of the two variables, giving a continuum of ENZ points.
The tuning of the ENZ point across a range of frequencies is achieved \textit{in~situ} by simply tilting the orientation of the plasmonic cavity relative to the magnetic field. 
Because of electron confinement in a 2DEG or MQW, only the normal component $B_\perp$ controls the cyclotron resonance and is responsible for tuning the ENZ and parallel permittivity components. From \eref{eq:axion-maxwell}, this means that the parallel component $B_\parallel$ is mainly responsible for and directly proportional to the boosted axion-induced current and can theoretically be as large as possible at each resonance frequency in order to maximize the signal. The independent control in tuning and signal enhancement with two magnetic field components allows the \SQWARE{} to rapidly scan across axion masses efficiently in the meV regime without complex mechanical movement or structural modification as in the Magnetized Disk and Mirror Axion Experiment (MADMAX)~\cite{millar_dielectric_2017, group_dielectric_2017} nor be constrained in frequency range by the strength of the magnetic field responsible for axion-photon conversion as in TOpolOgical Resonant Axion Detection (TOORAD)~\cite{li_dynamical_2010,marsh_proposal_2019,schutte-engel_axion_2021, qiu_observation_2025}.  

The MQW approach is further enabled by advances in material growth and thick wafer-scale fabrication, which now allow for high-purity, low-loss structures with engineered electron densities and barrier configurations \cite{li_mbe_2015,zhang_collective_2016,chung_ultra-high-quality_2021, chung_understanding_2022, hale_multi-mode_2025,li_multi-watt_2017, sharma_study_2025,brandstetter_thz_2012}.
For the frequency range of interest, the MQW structure behaves as a plasmonic cavity with highly tunable resonance and low dissipation, verified both analytically and numerically; see simulation details in SM~\cite{supplement}. The optimization of the boost and resonance quality, and chosen material parameters, is central to achieving the projected \SQWARE{} sensitivity.

\textit{Design Modularity}---We demonstrate the prospect of a conventional MQW forming a single magnetoplasmonic cavity to radiate photons in the presence of an axion field, tunable simply with the background magnetic field. Multiple MQWs in series with optimal spacing may also be designed for an enhanced coherent signal strength, as described in MADMAX~\cite{millar_dielectric_2017, group_dielectric_2017}. Additionally, more complex optimization of the structural design can be accomplished with machine learning, especially if the MQWs require thousands of layers, each fabricated with unique thickness, doping and electro-optical properties to enhance the axion-induced signal~\cite{jung_simulation_2022, mcdonald_scanning_2022}. The unique adaptability of the \SQWARE{} provides the advantage for labs with varying growth capabilities to target new mass ranges by designing their own MQWs using GaAs/AlGaAs, InGaAs/InP, GaN/AlGaN, or other magnetoplasmonic systems based on material availability and optimize structural parameters based on magnetic field limitations and desired axion targets. For example, smaller-scale MQWs with fewer layers can target shorter wavelengths, and lower magnetic fields are typically sensitive to higher axion masses.

\textit{Example \SQWARE{}}---We consider three example configurations to specifically target the $\sim1$~meV mass range, which correspond to currently available components for Config.~1 or high mass (HM) targeting $2$--$5$\,meV masses, near-future improvements for Config.~2 or low mass (LM) targeting $0.6$--$2$\,meV masses with thicker MQW cavities and more efficient photodetectors, and more ambitious, longer-term advances for Config.~3 or Low Mass Sensitive (LM-S) also targeting $0.6$--$2$\,meV masses with even thicker, larger, and high-quality MQWs, more sensitive photodetectors, and stronger magnets, achievable with anticipated technological advances in the next decade.   
Key parameters are accessible in \tref{parameterstable}; see more details in Sec.~III in SM~\cite{supplement}.
\begin{table}[h]
\centering
\begin{tabular}{l c c c} 
 \hline
Parameter& \,\,\,\,\,\,\,Config.\,1\,\,\,\,\,\,&\,\,\,\,\,\,\,Config.\,2\,\,\,\,\,\,&\,\,\,\,\,\,\,Config.\,3\,\,\,\,\,\,\\ 
 & (HM)&(LM)&(LM-S)\\ 
 \hline
$D$ [mm] & $2$ & $10$ & $20$ \\
$W$ [cm] & $3$ & $3$ & $5$ \\
$d_\text{Barrier}$ [nm] & $90$ & $150$ & $90$ \\
$B$ [T] & $36$ & $36$ & $50$ \\
$\tau$ [ns] & $1.7$ & $1.7$ & $4$\\
$\Gamma_{\text{dark}}$ [mHz] & $1$ & $1$ & $0.1$ \\
$\eta$ [\%]& $7$& $20$& $35$\\
$t_\text{obs}$ [min]& $84$& $34$& $7$\\
\hline
\end{tabular}
\caption{Summary of parameters for three example benchmark \SQWARE{} configurations: HM (high Mass), LM (low Mass), and LM-S (low mass sensitive). For all example scenarios, the temperature $T=0.3$~K, electron density $n_\mathrm{e} = 3 \times 10^{11}$~cm$^{-2}$, quantum well width $d_\text{QW} = 30$~nm, and total observation time $t_\text{tot} = 300$~d. As MQW cavities get larger and scattering times increase, larger signal boosts are possible at lower frequencies and sharper resonances result in shorter integration times.}
\label{parameterstable}
\end{table}

\textit{MQW structure and geometry}---The square-shaped MQW heterostructure may be fabricated by MBE or similar growth processes. The total thickness $D$ and width $W$ set the plasmonic volume: larger areas linearly enhance the signal power, while the total layer thickness is optimized for ENZ resonance in the desired axion mass range. See Sec.~II~C in SM~\cite{supplement} for effects of finite area on boost factor and Sec.~II~D in SM~\cite{supplement} for effects of finite area under an axion field with finite axion coherence length~\cite{supplement}. The axion field is modeled using~\rref{amaral_vector_2024}.
Each example MQW configuration uses a GaAs/AlGaAs system and fixes $d_{\text{QW}} = 30$~nm and $n_\mathrm{e} = 3\times10^{11}$~cm$^{-2}$, while discretely varying $D$, $W$, and $\tau$ according to technological feasibility, and optimizing $d_\text{Barrier}$ to $90$~nm for Configs.~1 and 3 and $150$~nm for Config.~2.
Standard fabrication of MQWs (in MBE) yields up to $W = 5$~cm, used in Config.~3, while a more conservative value of $W=3$~cm is used for Configs.~1 and 2, to fit in the small sample space of currently available magnets.
Configuration\,$1$ utilizes a $D=2$~mm thick MQW and targets the $2$--$5$\,meV axion mass range. Configurations~$2$ and $3$ use thicker MQWs with $D=10$ and $D=20$~mm, respectively, allowing for smaller permittivities and larger signal boosts at lower frequencies, i.e., $0.6$--$2$~meV mass range. Configurations\,1 and 2 utilize experimentally obtained values for mobility ($\mu_\mathrm{e} = 44\times10^6$~cm$^2$/Vs~\cite{chung_ultra-high-quality_2021}), corresponding to $\tau = 1.7$~ns for a GaAs 2DEG (for which $m^* \approx 0.067m_\mathrm{e}$, where $m_\mathrm{e}$ is the electron mass), while Config.\,3 uses theoretically achievable values~\cite{chung_understanding_2022, hwang_limit_2008}, $\mu_\mathrm{e} = 100\times10^6$\,cm$^2$/Vs or $\tau = 4$\,ns, for lower loss and better sensitivity. See Sec.~II~F in SM~\cite{supplement} for the effects of expected electron density nonuniformity in the 2DEGs~\cite{szerling_mid-infrared_2009, saito_growth_1987, gardner_modified_2016, manfra_molecular_2014,chung_spatial_2019} on the boost factor~\cite{supplement}.

\textit{Magnetic field}---A strong uniform dc magnetic field $\boldsymbol{B}$ is crucial both for maximizing axion-photon conversion and for tuning the ENZ resonance. 
A key requirement for the \SQWARE{} is magnetic field homogeneity across the plasmonic cavity. The ENZ point typically exhibits sharp resonance, depending on the material quality, making the signal highly sensitive to variations in $\boldsymbol{B}$. Inhomogeneities of even a hundred ppm across the sample can broaden or shift the resonance, reducing the effective signal and degrading sensitivity (see Sec.~II~E in SM~\cite{supplement}). 
To scan axion masses with zero mechanical movement, the magnet could be dynamically tuned using the current, with a fixed sample orientation, or using vector magnets~\cite{hadjigeorgiou_vector_2021}, but this may sacrifice the sensitivity at higher masses, which would require decreasing the entire magnetic field strength for tuning and neglects the available strength in the magnetic field for axion-photon production instead.
The \SQWARE{} example configurations assume a rotatable sample stage in a fixed high-field magnet, allowing the sample to tilt such that $B_\perp$ and $B_\parallel$ can be adjusted for efficient mass scanning. 
Both Configs.\,1 and 2 utilize the 36-T NMR magnet at the National High Magnetic Field Laboratory, which boasts a 1-ppm inhomogeneity over a 3-cm sample region~\cite{brandt_national_2001}, with an optional tilting probe available, while Config.~3 assumes a next-generation 5-cm, 50-T magnet to generate larger signals~\cite{noauthor_opportunities_2005}. Future studies may be done on the use of weaker magnets and a combination of higher electron density wells ($n_\mathrm{e} \gtrsim 10\times10^{11}$~cm$^{-2}$) and thinner barriers ($d_\text{Barrier} \lesssim 60$~nm) to target the same mass range, albeit weaker sensitivity.

\textit{Operation and scanning}---Mass scanning can be easily performed by varying the angle $\theta$ between the surface normal of the magnetoplasmonic cavity and the magnetic field, thereby shifting the ENZ point and resonance frequency to optimize the boost. The overall maximum boost for each example configuration is $\beta_\text{max}=70$, $115$, and $390$, which occur at masses $m_\mathrm{a} = 5.0$, $1.6$, and $1.1$~meV, corresponding to angles $\theta = 86^\circ$, $72^\circ$, and $60^\circ$, respectively. The scan rate is determined by the full-width-half-maximum or quality factor $Q$ of the boost factor as a function of frequency at a fixed angle. Several axion masses within a frequency band can be searched simultaneously when the detector's minimum resonance width, set by the maximum quality factor, which is $Q_\mathrm{max} = 6\times10^3$, $1\times10^4$, and $5\times10^4$ for the example configurations respectively, is wider than the cold DM axion linewidth, which has quality factor $\mathcal{O}(10^{6})$~\cite{adams_axion_2023}. With optimized MQW geometry and magnetic field control, the experiment can efficiently scan large regions of parameter space in a practical runtime, for total measurement scan time $t_\text{tot}$ and integration time per resonance frequency $t_\text{obs}$ (see more details on scanning in Sec.~I~F in SM~\cite{supplement}). The scan protocol can be adapted to prioritize regions of greatest theoretical or experimental interest. The experiment can be dynamically calibrated using ellipsometry or similar reflection or transmission spectroscopy measurement to account for drifts and shifted resonances as well as boost and quality factors~\cite{chen_introduction_2022,kriisa_cyclotron_2019}. 

\textit{Lens and photodetector}---Signal photons emitted from the magnetoplasmonic cavity are focused with ideal efficiency simply set to $100\%$, as parabolic mirrors or metalens can reach focus efficiencies above $90\%$ realistically~\cite{legaria_highly_2021, yang_high_2023,chen_3d-printed_2025,lindlein_focusing_2019, fleming_blazed_1997}. The photons are then measured with a THz photodetector, whose dark count and quantum efficiency directly set the experiment’s sensitivity. Quantum dot detectors offer state-of-the-art quantum efficiency $\eta$ and low dark count rates $\Gamma_{\text{dark}}$ in the $\sim\mathcal{O}(1)$~meV range~\cite{komiyama_single-photon_2000,astafiev_single-photon_2002,komiyama_single-photon_2011,kajihara_terahertz_2013,shaikhaidarov_detection_2016}. Configurations~1 and 2 assume $\eta \sim 0.07$ (shown experimentally) and $\eta \sim 0.2$ (with better antenna coupler), respectively, both with $\Gamma_{\text{dark}} \sim 1$~mHz~\cite{komiyama_single-photon_2011}, while Config.~3 assumes $\eta \sim 0.35$ (shown experimentally for higher frequencies)~\cite{nakai_development_2024}, with $\Gamma_{\text{dark}} \sim 0.1$~mHz, which may be feasible in the future, as theoretically expected dark counts reach $\upmu$Hz for next-generation quantum dot detectors~\cite{komiyama_single-photon_2011}. 

\textit{Sensitivity}---The projected example \SQWARE{} sensitivity to axion DM is set by requiring a 95\% confidence exclusion. In the background-dominant scenario, the expected axion signal must exceed twice the standard deviation of detector noise, approximated from Poisson statistics. The dominant noise is the photodetector’s dark count rate $\Gamma_\text{dark}$, with other sources (e.g., thermal photons or cosmic rays) rendered negligible by sub-Kelvin operation and sufficient shielding. The minimum detectable signal rate is thus $\Gamma_\text{signal} >2\sqrt{\Gamma_{\text{dark}} / t_\text{obs}}$, 
where $t_\text{obs}$ is the integration time~\cite{schutte-engel_axion_2021,bityukov_new_1998}. 
In background-free scenarios (e.g., with a next-generation detector with dark count less than 1 per $t_\text{obs}$), a threshold of three detected photons is required~\cite{liu_broadband_2022, navas_review_2024}. Assuming ideal focusing, the signal rate detected by a photodetector with quantum efficiency $\eta$, boost factor $\beta$, magnetic field $|\boldsymbol{B}|$ , and plasmonic cavity area $W^2$ is $\Gamma_\text{signal} =  \eta|\boldsymbol{E}|^2 W^2 / 2\omega$, where $|\boldsymbol{E}| = \beta \gagg |\boldsymbol{B}| a_0$. 
The sensitivity to the axion-photon coupling in the background-dominated regime, over integration time, $t_\mathrm{obs}$, is therefore
\begin{align}\label{eq:gagg_sensitivity}
    \gagg > \,\,
5\times 10^{-14}\, \text{GeV}^{-1}  \left[ \frac{36\, \text{T}}{|\boldsymbol{B}|} \right]  \left[ \frac{100}{\beta(m_\mathrm{a})} \right]\left[ \frac{20\%}{\eta} \right]^{\frac{1}{2}}&\\
\times  \left[ \frac{9 \,\text{cm}^2}{W^2} \right]^{\frac{1}{2}} \left[ \frac{\Gamma_{\text{dark}}}{1 \,\text{mHz}} \right]^{\frac{1}{4}} \left[ \frac{m_\mathrm{a}}{1\, \text{meV}} \right]^{\frac{3}{2}} \left[ \frac{30 \, \text{days}}{t_\text{obs}} \right]^{\frac{1}{4}}&
\;.\nonumber
\end{align} 
The boost factor $\beta(\omega,\theta) = \beta(m_\mathrm{a})$ depends on the axion mass $m_\mathrm{a}$ implicitly through the resonance frequency $\omega = m_\mathrm{a}$ and angle $\theta$. A similar formula can be derived for the background-free regime, which applies to Config.~3 for wide scans. 

\begin{figure}[h!]
    \centering
    \includegraphics[width=1\linewidth]{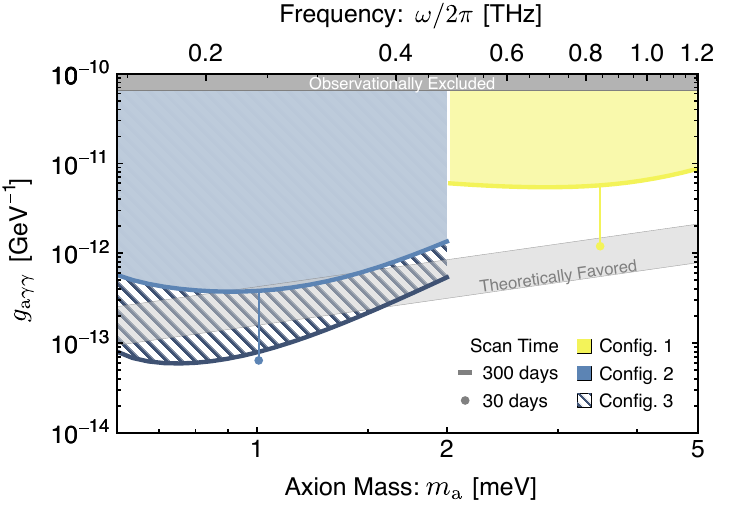}
\caption{\label{fig:parameter_space}
    Projected sensitivity of an example \SQWARE{} to the axion mass $m_\mathrm{a}$ and axion-photon coupling $\gagg$ for three configurations (in yellow, blue, and striped navy) for $t_\text{tot} =$ $300$-d wide mass sweeps or $30$-d single resonance runs, generating significant coverage of the axion parameter space beyond current experimental constraints, with CAST limits in the top gray band~\cite{adams_axion_2023}. For optimistic configurations, theoretically favorable QCD axion models may be probed, with Dine-Fischler-Srednicki-Zhitnitsky and Kim-Shifman-Vainshtein-Zakharov benchmarks~\cite{luzio_landscape_2020} in the diagonal light gray band.
    }
\end{figure}
In Fig.~\ref{fig:parameter_space}, we show the projected sensitivity of an example \SQWARE{} for the three benchmark configurations provided in \tref{parameterstable}. 
For narrow scans, we assume a total exposure time of $t_\text{tot} = 30$~d, where the detector sits at a fixed angle and resonance frequency, portrayed as narrow lines at $m_\mathrm{a} = 1$ and $3.5$~meV for Configs.~1 and 2, respectively. 
For wide scans, we assume a $t_\text{tot} = 300$~d sweep over a range of axion masses corresponding to several thousand scans, each with integration time $t_\text{obs} = 84$, $34$, and $7$~min, for the three example configurations, respectively.
The conservative and realistic designs of Configs.~1 and 2 are already projected to achieve leading limits, surpassing the CERN Axion Solar Telescope (CAST) helioscope for axion masses $0.6$--$5$~meV, and the aggressive Config.~3 can probe the favored QCD axion band at axion masses of $0.6$--$2$~meV~\cite{adams_axion_2023}. The \SQWARE{} can simultaneously probe dark photon DM, which does not require a magnetic field for photon-dark photon conversion~\cite{holdom_two_1986, fabbrichesi_dark_2021}; see End Matter for projected sensitivities of the three configurations. 

\textit{Conclusion}---We have introduced the \SQWARE{}, a new approach for direct axion DM detection at meV masses, leveraging recent advances in quantum semiconductor heterostructures to create high-quality magnetoplasmonic cavities with ENZ response and simple tuning mechanisms. This platform offers a feasible path for experimental laboratories to design and probe the unexplored axion parameter space, including the QCD axion band, with near-term technology. Experiment design details, extended calculations, and simulations are provided in SM~\cite{supplement}.\\

\textit{Acknowledgments}---This material is based upon work supported (in part, K.~S. and T.~X.) by the  National Science Foundation under Grant No.~PHY-2514896, (in part, A.~J.~L.) by the National Science Foundation under Grant No.~PHY-2412797, (in part, J.~K.) by the U.S. Army Research Office under Awards No.~W911NF-21-1-0157 and No.~W911NF-25-2-0150, the Gordon and Betty Moore Foundation under Grant No.~11520, and the Robert A. Welch Foundation under Grant No.~C-1509, (in part, J.~M.) by the National Science Foundation Graduate Research Fellowship under Grant No.~1842494, and (in part, J.~M. and S.~H.) by the National Science Foundation under Grants No.~ECCS-2246564 and No.~ECCS-1943895, the Air Force Office of Scientific Research (AFOSR) under Grant No.~FA9550-22-1-0408, the Robert A. Welch Foundation under Grant No.~C-2144, and Office of Naval Research
(ONR) under Grant No. N000142512387. Thanks to Mudit Jain for code generating the axion field. We thank Alexey Belyanin, Mustafa Amin, Dorian Amaral, and R.C. Woods for helpful discussions. Thanks to Haaniya Mehrani for the \SQWARE{} logo.\\

\textit{Data availability}---The data that support the findings of this article are not publicly available upon publication because it is not technically feasible and/or the cost of preparing, depositing, and hosting the data would be prohibitive within the terms of this research project. The data are available from the authors upon reasonable request.


    \let\oldaddcontentsline\addcontentsline
    \renewcommand{\addcontentsline}[3]{}
    \bibliographystyle{apsrev4-2}

    \let\addcontentsline\oldaddcontentsline

\newpage
\setcounter{equation}{0}
\setcounter{table}{0}
\setcounter{secnumdepth}{2}
\makeatletter
\renewcommand{\theequation}{B\arabic{equation}}
\renewcommand{\thetable}{B\arabic{table}}

\onecolumngrid

\begin{center}
    \large{\textbf{End Matter}}
\end{center}
\twocolumngrid

\textit{Validation of effective medium theory}---The analytical boost factor from \eref{boostonedisk} assumes an averaged or effective permittivity, given by \eref{eq:effmedperm}, under the approximations of effective medium theory (EMT). To validate EMT for an example \SQWARE{}, the discrete layers of GaAs quantum wells (QWs) and AlGaAs barriers, with permittivities $\varepsilon_\text{QW}$ and $\varepsilon_\text{Barrier}$ respectively, are simulated in a 1D \textsc{comsol} simulation, (see Sec.~II~A in SM~\cite{supplement} for details, including how the axion field can be modeled following MADMAX's technique~\cite{Gardikiotis2020_presentation}). The numerically solved electric field enhancement is compared with the analytical expression of boost factor as a function of the effective permittivity. The two individual layer thicknesses $d_\text{QW}$ and $d_\text{Barrier}$ are varied for two cases, while the total thickness $D$ and the ratio between the two are kept constant. The imaginary part of the permittivity $\varepsilon''_\text{QW}$ is fixed while the real part $\varepsilon'_\text{QW}$ is tuned to locate the maximum boost.

\begin{figure}[h]
\centering
\includegraphics[width=0.475\textwidth]{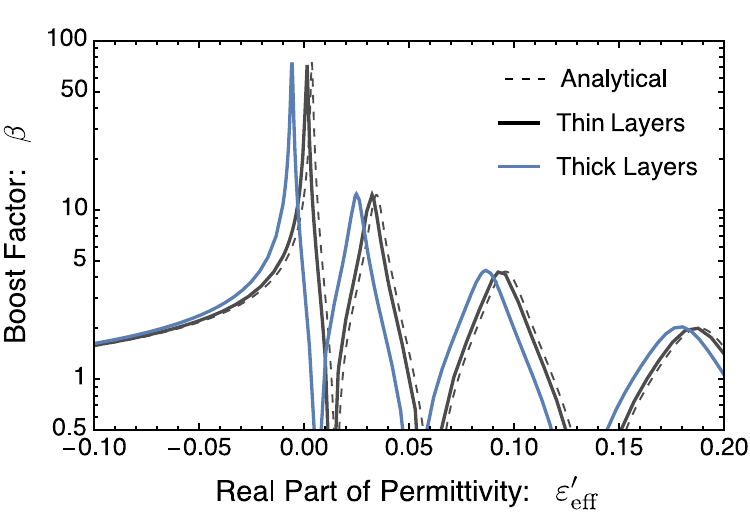}
\caption{Validation of EMT for an example MQW structure in \textsc{comsol}. The dashed black line shows the analytical boost factor as a function of the real part of the effective permittivity $\varepsilon_\text{eff}'$, while the imaginary part is $\varepsilon_\text{eff}''$ is fixed at $2 \times 10^{-4}$, while the black and blue solid lines show numerical results for thin layers, where $d_\text{Barrier}=0.0025\lambda_\text{vac}$ and $d_\text{QW}=0.0005\lambda_\text{vac}$, and thicker layers where $d_\text{Barrier}=0.005\lambda_\text{vac}$ and $d_\text{QW}=0.001\lambda_\text{vac}$, respectively, both with total thickness $D=1$\,cm. Thinner layers emulate EMT; thicker layers break down EMT and shift the resonance toward negative permittivities, with peak boost factor preserved.}
\label{emtfigurem}
\end{figure}

As shown in \fref{emtfigurem}, the analytical boost factor in \eref{boostonedisk} closely matches the results when individual layers are much thinner than the in-medium wavelength, confirming the validity of EMT. For thicker layers, EMT breaks down, but the maximum achievable boost and shape of the boost factor vs permittivity curves are maintained (realized at a more negative real part of the permittivity), further illustrated in Sec.~II~B in SM~\cite{supplement}. For all MQW structures in the example \SQWARE{} configurations, EMT is an excellent approximation, as the layer thicknesses are on the order of tens of nm, much smaller than the in-medium wavelengths on the order of hundreds of $\upmu$m.

\textit{Sensitivity of \texttt{SQWARE} to dark photon dark matter}---Dark photons may interact with ordinary electromagnetic photons through a kinetic mixing, which is parametrized by a dimensionless mixing parameter $\chi$~\cite{holdom_two_1986}.  
Unlike axion-photon conversion, dark photon-photon conversion occurs spontaneously without a magnetic field. The \SQWARE{}, like many axion search strategies, can search for dark photon DM at the same time as it searches for axion DM~\cite{fabbrichesi_dark_2021}.  
As the plasmonic cavity is rotated in a magnetic field, the resonance frequency varies, scanning across DM masses.  
Alternatively, one could optimize the search for dark photon DM by holding the cavity fixed, with the magnetic field normal to its surface, and scan across dark photon masses by varying the field strength.

We estimate the projected sensitivity of an example \SQWARE{} to dark photon DM following~\rref{liu_broadband_2022}.
The formula for axion DM in \eref{eq:gagg_sensitivity} is modified with the replacements
\begin{subequations}
\begin{align}
    \gagg B_\parallel \rightarrow m_{A^\prime} \chi 
    ,\\
    m_\mathrm{a} \to m_{A^\prime} 
    ,\\
    a_0 \to a_0 / \sqrt{3} 
    \;,
\end{align}
\end{subequations}
where $m_{A^\prime}$ is the dark photon mass, and the factor of $\sqrt{3}$ accounts for the dark photon's random polarization.
The dark photon sensitivity is
\begin{equation}
\begin{aligned}
    \chi  > 
    6\times 10^{-16} 
    \left[ \frac{100}{\beta(m_{A^\prime})} \right]^{} 
    \left[ \frac{0.2}{\eta} \right]^{\frac{1}{2}}& \\
    \times \left[ \frac{9~\text{cm}^2}{w^2} \right]^{\frac{1}{2}}
     \left[ \frac{\Gamma_\text{dark}}{1~\text{mHz}} \right]^{\frac{1}{4}} 
    \left[ \frac{m_{A^\prime}}{1~ \text{meV}} \right]^{\frac{1}{2}} 
    \left[ \frac{30 ~ \text{d}}{t_\text{obs}} \right]^{\frac{1}{4}}&.
\end{aligned} 
\end{equation}

\begin{figure}[h]
    \centering
    \includegraphics[width=1\linewidth]{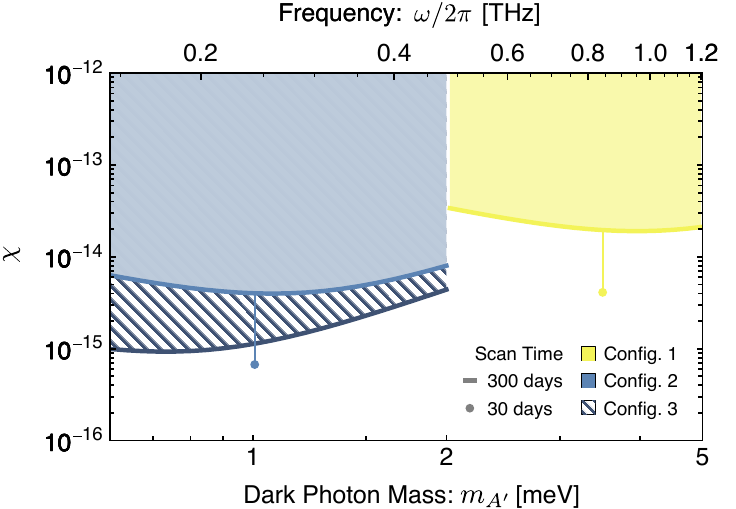}
    \caption{\label{fig:chi_sensitivity}
    Projected sensitivity of an example \SQWARE{} to the dark photon mass $m_{A'}$ and dark photon-photon mixing $\chi$ for three configurations (in yellow, blue, and striped navy) for $t_\text{tot} =$ $300$-d wide mass sweeps or $30$-d single resonance runs.}
\end{figure}

In \fref{fig:chi_sensitivity} we show the projected sensitivity of an example \SQWARE{} to dark photon DM for the three design configurations in \tref{parameterstable}.
By comparing data from runs optimized for axion sensitivity, the \SQWARE{} can place competitive limits on dark photon DM in the meV mass range simultaneously.

\onecolumngrid
\newpage

\renewcommand{\SQWARE}[1]{\texttt{SQWARE}}

\onecolumngrid
\begin{center}
\textbf{\large 
Supplemental Material for ``Quantum Semiconductor Heterostructures for meV Axion Dark Matter Detection''
}
\end{center}
\setcounter{equation}{0}
\setcounter{figure}{0}
\setcounter{table}{0}

\setcounter{secnumdepth}{2}
\makeatletter
\onecolumngrid
\renewcommand{\theequation}{S\arabic{equation}}
\renewcommand{\thefigure}{S\arabic{figure}}
\renewcommand{\thetable}{S\arabic{table}}
\thispagestyle{empty}
This Supplemental Material provides additional details and results for the analyses discussed in the main Letter. 

\tableofcontents
\pagenumbering{arabic}

\section{Theory of axion-induced radiation from plasmonic cavities}
\label{sec:theory}

In this section, we discuss how the axion field and the electromagnetic field interact with one another, how a plasmonic material responds to axion dark matter (DM) in an external magnetic field, how the electric field strength is enhanced at an epsilon-near-zero (ENZ) point, how a quantum well forming a two-dimensional electron gas (2DEG) provides a magnetically-tunable ENZ point, how a stack of multiple quantum wells (MQWs) behaves as a plasmonic cavity, and how radiation from a plasmonic cavity is calculated. 

\subsection{Axion electrodynamics in a medium}
\label{sub:axion_electro}

Interactions between an axion and electromagnetism are governed by the Lagrangian 
\begin{equation}
    \mathcal{L} 
    = -\tfrac{1}{4} F_{\mu \nu} F^{\mu \nu} 
    - J^\mu A_\mu 
    + \tfrac{1}{2} \partial_\mu a \partial^\mu a 
    - \tfrac{1}{2} m_\mathrm{a}^2 a^2 
    - \tfrac{1}{4} \gagg a F_{\mu \nu} \tilde{F}^{\mu \nu} 
    \;,
\end{equation}
where $a(x)$ is the axion field, $A_\mu(x)$ is the electromagnetic potential, $F_{\mu\nu}(x)$ is the electromagnetic field strength tensor, $\tilde{F}^{\mu\nu}(x)$ is its dual, $J^\mu(x)$ is the electromagnetic current density, $m_\mathrm{a}$ is the axion mass, and $\gagg$ is the axion-photon coupling. 
We follow steps derived in~\rref{supplemental_millar_dielectric_2017} and work in Heaviside-Lorentz units and remove factors of $\hbar$ and $c$.  
The field equations and electromagnetic Bianchi identity are 
\begin{subequations}
\begin{align}
    (\partial_\mu \partial^\mu + m_\mathrm{a}^2) a 
    & = - \tfrac{1}{4} \gagg F_{\mu\nu} \tilde{F}^{\mu\nu} \;, \\
    \partial_\mu F^{\mu\nu} 
    & = J^\nu - \gagg \tilde{F}^{\mu\nu} \partial_\mu a \;, \\
    0 & = \partial_{\lambda} F_{\mu\nu} + \partial_{\mu} F_{\nu\lambda} + \partial_{\nu} F_{\lambda\mu} \;.
\end{align}
\end{subequations}
The in-medium form of Maxwell’s equations is modified by the axion to 
\begin{subequations}\label{eq:Maxwell}
\begin{align}
    \ddot{a} - \nabla^2 a + m_\mathrm{a}^2 a & = \gagg \Evec \cdot \Bvec \;, \\ 
    \dvec \cdot \Dvec - \rho_\mathrm{free} & = - \gagg \Bvec \cdot \dvec a \;, \\
    \dvec \times \Hvec - \dot{\Dvec}-\Jvec_\mathrm{free} & = \gagg \Bvec \dot{a} - \gagg \Evec \times \dvec a \;, \\
    \dvec \cdot \Bvec & = 0 \;, \\
    \dvec \times \Evec + \dot{\Bvec} & = 0
    \;.
\end{align}
\end{subequations}
Here, $\Evec(\rvec,t)$, $\Bvec(\rvec,t)$, $\Dvec(\rvec,t)$, and $\Hvec(\rvec,t)$ denote the electric field, magnetic field, displacement field, and magnetizing field; additionally, $\rho_\mathrm{free}(\rvec,t)$ and $\Jvec_\mathrm{free}(\rvec,t)$ denote the free charge and current densities. 
The fields admit the Fourier representations:  
\begin{align}
    X(\rvec,t) = \int_{-\infty}^{\infty} \! \frac{d\omega}{2\pi} \int \! \! \frac{d^3\kvec}{(2\pi)^3} \, \hat{X}(\kvec,\omega) \, e^{-i \omega t + i \kvec \cdot \rvec} 
    \quad \text{for $X = a, \Evec, \Bvec, \Dvec, \Hvec, \rho_\mathrm{free}$, and $\Jvec_\mathrm{free}$}
    \;,
\end{align}
where $\kvec$ is the wave vector, $k = |\kvec|$ is the corresponding wavenumber, and $\omega$ is the angular frequency.  
In Fourier space, these equations take the form 
\begin{subequations}\label{eq:Maxwell_Fourier}
\begin{align}
    \bigl( \omega^2 - k^2 - m_\mathrm{a}^2 \bigr) a & = - \gagg \hat{\Evec} \otimes \hat{\Bvec} \;, \\ 
    \kvec \cdot \hat{\Dvec} - \hat{\rho}_\mathrm{free} & = -\gagg \hat{\Bvec} \otimes \kvec^\prime \hat{a} \;, \\
    \label{eq:Ampere_Law}
    \kvec \times \hat{\Hvec} + \omega  \hat{\Dvec} + i \hat{\Jvec}_\mathrm{free} & = - \gagg \hat{\Bvec} \otimes \omega^\prime \hat{a} - \gagg \hat{\Evec} \otimes \kvec^\prime \hat{a} \;, \\
    \kvec \cdot \hat{\Bvec} & = 0 \;, \\
    \kvec \times \hat{\Evec} - \omega \hat{\Bvec} & = 0 \;, 
\end{align}
\end{subequations}
where $\otimes$ denotes convolutions over $\omega^\prime$ and $\kvec^\prime$. 
Whereas electromagnetism has linear field equations, axion-electromagnetism has nonlinear field equations due to the terms proportional to the axion-photon coupling $\gagg$.  

We model the response of the medium with the linear constitutive relations 
\begin{subequations}\label{eq:constitutive_relation}
\begin{align}
    \hat{\Dvec} = \bigl( \mathbb{I} + \hat{\chi} \bigr) \, \hat{\Evec} 
    \;, \qquad 
    \hat{\Hvec} = \hat{\mu}^{-1} \, \hat{\Bvec} 
    \;, \qquad 
    \hat{\rho}_\mathrm{free} = 0 
    \;, \quad \text{and} \qquad 
    \hat{\Jvec}_\mathrm{free} = \hat{\sigma} \, \hat{\Evec} 
    \;,
\end{align}
where $\hat{\chi}(\kvec,\omega)$ is the electric susceptibility tensor, $\hat{\mu}^{-1}(\kvec,\omega)$ is the inverse magnetic permeability tensor, $\hat{\sigma}(\kvec,\omega)$ is the electric conductivity tensor, and $\mathbb{I}$ is the identity matrix. .  
In general, these tensors may be anisotropic and complex.  
The terms in \eref{eq:Ampere_Law} are written as 
\begin{align}
    \omega \hat{\Dvec} + i \hat{\Jvec}_\mathrm{free} = \omega \hat{\varepsilon} \hat{\Evec} 
    \qquad \text{with} \qquad 
    \hat{\varepsilon} = \mathbb{I} + \hat{\chi} + \frac{i}{\omega} \, \hat{\sigma} 
    \;,
\end{align}
\end{subequations}
where $\hat{\varepsilon}(\kvec,\omega)$ is the electric permittivity tensor.
Expressions for $\hat{\varepsilon}(\kvec,\omega)$ may be found in \sref{sub:cyclotron}. 

We are interested in the electromagnetic radiation that results when a volume of space containing axion DM is exposed to a magnetic field.  
We model the applied external magnetic field as a static and homogeneous background $\Bvec_\mathrm{ext}$.  
This is a reasonable approximation for \SQWARE{} provided that field homogeneity can be maintained over the scale of the plasmonic cavity. 
Allowing for perturbations on top of this background, we write the magnetic field as $\Bvec(\rvec,t) = \Bvec_\mathrm{ext} + \Bvec_\mathrm{prop}(\rvec,t)$, and the corresponding Fourier transform is $\hat{\Bvec}(\kvec,\omega) = \Bvec_\mathrm{ext} \, (2\pi)^4 \delta(\omega) \, \delta(\kvec) + \hat{\Bvec}_\mathrm{prop}(\kvec,\omega)$.

We model the axion DM as a harmonically oscillating and spatially homogeneous background $a_0 \cos(m_\mathrm{a} t + \varphi_\mathrm{a})$, which oscillates with angular frequency $\omega = m_\mathrm{a}$, amplitude $a_0$, and phase $\varphi_\mathrm{a}$. 
Assuming that the axion makes up all the DM ($\rho_\mathrm{a} = \rho_\mathrm{dm}$), the amplitude is related to the local DM energy density $\rho_\mathrm{dm} \approx 0.3$\,GeV/cm$^3$ through $a_0 = \sqrt{2 \rho_\mathrm{a}} / m_\mathrm{a} \approx 1.52 \, \mathrm{eV}\, (m_\mathrm{a} / \mathrm{meV})^{-1}$.  
A more accurate modeling would allow both the amplitude and phase of this field to vary stochastically in space with coherence length $l_\mathrm{coh} = 2\pi / m_\mathrm{a} v_\mathrm{a} \approx 42 \, \mathrm{cm}\, (m_\mathrm{a} /  \mathrm{meV})^{-1} (v_\mathrm{a} / 220 \, \mathrm{km}/\mathrm{sec})^{-1}$ and to vary stochastically in time with coherence time $t_\mathrm{coh} = 2\pi / m_\mathrm{a} v_\mathrm{a}^2 \approx 1.9 \, \mu\mathrm{s}\, (m_\mathrm{a} /  \mathrm{meV})^{-1} (v_\mathrm{a} / 220 \, \mathrm{km}/\mathrm{sec})^{-2}$, where $v_\mathrm{a} \approx 220 \, \mathrm{km}/\mathrm{sec}$ is the local DM velocity dispersion.  
However, for the parameters of interest, the coherence length is much larger than the size of the plasmonic cavity in \SQWARE{} and the coherence time is much longer than the time scale for electromagnetic radiation; therefore, it is a good approximation to treat the axion field as homogeneous with fixed amplitude and phase. 
We validate this approximation using numerical simulation, and those results are reported in \sref{sub:axionhom}.  
Allowing for perturbations on top of this background, we write the axion field as $a(\rvec,t) = a_0 \, e^{-i m_\mathrm{a} t} + a_\mathrm{prop}(\rvec,t)$ and the corresponding Fourier transform is $\hat{a}(\kvec,\omega) = a_0 \, (2\pi)^4 \delta(\omega - m_\mathrm{a}) \delta(\kvec) + \hat{a}_\mathrm{prop}(\kvec,\omega)$.  
For intermediate calculations, we treat $a(\rvec,t)$ as a complex field, and its real part gives the observable axion field.  
Although the field equations of axion electrodynamics are nonlinear \eqref{eq:Maxwell}, this approach is justified because we study perturbations on the real $a(\rvec,t)$ and $\Bvec(\rvec,t)$ backgrounds, and we can neglect the nonlinear terms: $\Evec \cdot \Bvec_\mathrm{prop}$, $\Bvec_\mathrm{prop} \cdot \dvec a_\mathrm{prop}$, $\Bvec_\mathrm{prop} \dot{a}_\mathrm{prop}$, and $\Evec \times \dvec a_\mathrm{prop}$.  

By modeling the fields in this way, neglecting nonlinear terms, Maxwell's equations \eqref{eq:Maxwell_Fourier} simplify to 
\begin{subequations}\label{eq:Maxwell_axion}
\begin{align}
    \kvec \cdot \bigl( \mathbb{I} + \hat{\chi} \bigr) \hat{\Evec} & = 0 \;, \\
    \kvec \times \hat{\mu}^{-1} \hat{\Bvec} 
    + \omega \hat{\varepsilon} \, \hat{\Evec} 
    & = - \gagg \Bvec_\mathrm{ext} m_\mathrm{a} a_0 (2\pi)^4 \delta(\omega - m_\mathrm{a}) \delta(\kvec) \;, \\
    \kvec \cdot \hat{\Bvec} & = 0 \;, \\
    \kvec \times \hat{\Evec} - \omega \hat{\Bvec} & = 0 \;.
\end{align}
\end{subequations} 

Solutions of these equations can be written as 
\begin{align}\label{eq:prop_plus_ax}
    \hat{\Evec} = \hat{\Evec}_\mathrm{prop} + \hat{\Evec}_\mathrm{ax} 
    \qquad \text{and} \qquad 
    \hat{\Bvec} = \hat{\Bvec}_\mathrm{prop} + \hat{\Bvec}_\mathrm{ax} 
    \;,
\end{align}
where $\hat{\Evec}_\mathrm{prop}(\kvec,\omega)$ and $\hat{\Bvec}_\mathrm{prop}(\kvec,\omega)$ represent a general superposition of propagating electromagnetic waves, where 
\begin{subequations}\label{eq:Ea}
\begin{align}
    \hat{\Evec}_\mathrm{ax}(\kvec,\omega) = - \gagg \hat{\varepsilon}^{-1}(\zvec,m_\mathrm{a}) \Bvec_\mathrm{ext} a_0 (2\pi)^4 \delta(\omega - m_\mathrm{a}) \delta(\kvec)
    \qquad \text{and} \qquad 
    \hat{\Bvec}_\mathrm{ax}(\kvec,\omega) = \zvec 
    \;,
\end{align}
represent the axion-induced fields, and where they correspond to 
\begin{align} \label{eq:axion-induced electric field}
    \Evec_\mathrm{ax}(\rvec,t) = - \gagg a_0 \hat{\varepsilon}^{-1}(\zvec,m_\mathrm{a}) \Bvec_\mathrm{ext} \, e^{-i m_\mathrm{a} t} 
    \qquad \text{and} \qquad 
    \Bvec_\mathrm{ax}(\rvec,t) = \zvec 
    \;.
\end{align}
\end{subequations}
Note that $\hat{\varepsilon}^{-1}$ is the inverse of the permittivity tensor evaluated at $\kvec = \zvec$ and $\omega = m_\mathrm{a}$. 

If the entire system were in a vacuum, where $\hat{\varepsilon} = 1$, then the axion-induced electric field would be 
\begin{align}\label{eq:Evec_vacuum}
    \Evec_\mathrm{ax}(\rvec,t) 
    & = - \gagg a_0 \Bvec_\mathrm{ext} \, e^{-i m_\mathrm{a} t}  
    \qquad \text{(in vacuum)} 
    \;,
\end{align}
and it is useful to define $E_\mathrm{vac} = \gagg a_0 |\Bvec_\mathrm{ext}|$. 
However, the amplitude of the axion-induced electric field in a material, or in vacuum nearby to a material, can be much larger than $E_\mathrm{vac}$. 
To make a comparison between them, it is customary to define the dimensionless boost factor 
\begin{align}\label{eq:beta_def}
    \beta = \frac{\|\Evec(\rvec,t)\|}{E_\mathrm{vac}} 
    \;.
\end{align}
The total electric field $\Evec(\rvec,t) = \Evec_\mathrm{ax}(\rvec,t) + \Evec_\mathrm{prop}(\rvec,t)$ is the sum of the axion-induced electric field from \eref{eq:Ea} and an additional term arising from propagating or homogeneous solutions to the Maxwell equations.  
The double-bar notation $\|\Evec(\rvec,t)\| = \sqrt{\Evec^\ast \cdot \Evec}$ denotes both 3-vector norm and the complex modulus, and it gives the (real, positive) amplitude of the electric field.  
If the entire system were in a vacuum ($\hat{\varepsilon}=1$), the boost factor $\beta = 1$, but $\beta \gg 1$ can be realized in some materials. 
\subsection{Epsilon-near-zero point}
\label{sub:enz}
A large signal boost ($\beta \gg 1$) can be achieved in materials that admit epsilon-near-zero (ENZ) points.  
In this subsection, we discuss the general consequences of an ENZ point, and in the following subsection, we provide explicit materials that admit ENZ points. 
In an infinite, homogeneous, and isotropic material, the permittivity tensor may be written as $\hat{\varepsilon}(\kvec,\omega) = \varepsilon(\kvec,\omega) \, \mathbb{
I
}$ where $\varepsilon(\kvec,\omega)$ is the scalar permittivity.  
The axion-induced electric field from \eref{eq:Ea} simplifies to 
\begin{align}
    \Evec_\mathrm{ax}(\rvec,t) 
    & = - \gagg a_0 \frac{1}{\varepsilon(\zvec,m_\mathrm{a})} \Bvec_\mathrm{ext} \, e^{-i m_\mathrm{a} t}  
    \qquad \text{(isotropic material)} 
    \;,
\end{align}
and the corresponding boost factor is calculated using \eref{eq:beta_def} ($\Evec_\mathrm{prop} = 0$), which gives 
\begin{equation}\label{eq:beta_0}
    \beta 
    = \frac{1}{|\varepsilon(\zvec,m_\mathrm{a})|} 
    = \frac{1}{\sqrt{(\varepsilon^\prime)^2 + (\varepsilon^\pprime)^2}} 
    \qquad \text{(isotropic material)}
    \;.
\end{equation}

It is useful to denote the real and imaginary parts of the electric permittivity as $\varepsilon^\prime(\kvec,\omega) = \mathrm{Re}[\varepsilon]$ and $\varepsilon^\pprime(\kvec,\omega) = \mathrm{Im}[\varepsilon]$ such that $|\varepsilon| = [(\varepsilon^\prime)^2 + (\varepsilon^\pprime)^2]^{1/2}$.
Note that $\beta$ becomes arbitrarily large as $\varepsilon^\prime$ and $\varepsilon^\pprime$ both become arbitrarily small. 
In practice, there do not exist materials for which the electric permittivity has simultaneously vanishing real and imaginary parts.  
However, some materials admit special frequencies $\omega_\mathrm{ENZ}$, which are called ENZ points, at which $|\varepsilon|$ is close to zero, typically when $\varepsilon'$ crosses zero and $\varepsilon''$, proportional to loss, is small. 
If $\omega_\mathrm{ENZ}$ coincides with the axion mass $m_\mathrm{a}$ then the ENZ point provides for a strong enhancement of the axion-induced electric field. 
For \SQWARE{}, $\omega_\mathrm{ENZ}$ can be tuned using the external magnetic field, as we discuss in the following subsection.  
\subsection{Cyclotron resonance in a quantum well}
\label{sub:cyclotron}

In this subsection, we discuss three-dimensional and two-dimensional systems that develop ENZ points when exposed to a magnetic field.  
We discuss how two-dimensional quantum wells provide a platform for ENZ points that are tunable with the external magnetic field.  

For pedagogical purposes, we begin by discussing the electromagnetic response in a crystalline solid, which can be modeled as a three-dimensional electron gas (3DEG). We follow the derivation from~\rref{supplemental_hilton_cyclotron_2012, supplemental_li_vacuum_2018}. 
If a homogeneous and static magnetic field $\Bvec_\mathrm{ext}$ points in the $\evec_z$ direction ($\evec$ denotes unit 3-vector), then the conductivity tensor $\hat{\sigma}(\kvec,\omega)$ is anisotropic 
\begin{subequations}\label{eq:sigma_hat_linear}
\begin{align}
    \hat{\sigma} & = 
    \begin{pmatrix}
    \sigma_{x x} & \sigma_{x y} & 0 \\
    \sigma_{y x} & \sigma_{y y} & 0 \\
    0 & 0 & \sigma_{z z}
    \end{pmatrix}
\end{align}
where
\begin{align}
    \sigma_{xx} 
    = \sigma_{yy} 
    = \frac{\sigma_0 (1 - i \omega \tau)}{(1 - i \omega \tau)^2 + \omega_c^2 \tau^2} \;, 
    \qquad 
    \sigma_{xy} 
    = -\sigma_{yx} 
    = - \frac{\sigma_0\omega_c \tau}{(1 - i \omega \tau)^2 + \omega_c^2\tau^2} \;, 
    \qquad 
    \sigma_{zz} & =  \frac{\sigma_0}{1 + i \omega \tau} \;.
\end{align}
\end{subequations}
Here $\sigma_0 = n_\text{bulk}e\mu = n_\text{bulk}e^2\tau/m^*$ is the DC conductivity in the absence of a magnetic field, 
$n_\text{bulk}$ is the bulk electron number density, $e = \sqrt{4 \pi \alpha} \approx 0.303$ is the magnitude of the electron's electric charge, $\alpha \approx 1/137$ is the electromagnetic fine structure constant, $\tau$ is the characteristic electron scattering time, $m^* \approx 0.067 m_\mathrm{e} \approx 0.034$\,MeV is the electron effective mass in the GaAs layer 
\cite{supplemental_sze_physics_2021}, 
$\omega_c = e |\Bvec_\mathrm{ext}|/m^*$ is the electron cyclotron frequency, and $\mu = e\tau/m^*$ is the electron mobility.
Transforming to the circular polarization basis, $\evec_\text{CRA} = (\evec_x + i \evec_y) / \sqrt{2}$ and $\evec_\text{CRI} = (\evec_x - i \evec_y) / \sqrt{2}$ and $\evec_\perp = \evec_z$, the conductivity tensor diagonalizes: 
\begin{subequations}\label{eq:sigma_hat_circular}
\begin{equation}
    \hat{\sigma} = \mathrm{diag}(\sigma_\text{CRA},\ \sigma_\text{CRI},\ \sigma_\perp) \;,
\end{equation}
where 
\begin{align}
    \sigma_\text{CRA} 
    = \frac{\sigma_0}{1 - i (\omega - \omega_c) \tau} \;, 
    \qquad 
    \sigma_\text{CRI} 
    = \frac{\sigma_0}{1 - i (\omega + \omega_c) \tau} \;, 
    \qquad 
    \sigma_\perp = \frac{\sigma_0}{1 + i \omega \tau} \;.
\end{align}
\end{subequations}
The subscripts denote the cyclotron resonance active (CRA) and cyclotron resonance inactive (CRI) polarization modes. 
Note that $|\sigma_\text{CRA}|$ is maximized at the cyclotron resonance frequency $\omega = \omega_c$.  
The corresponding electric permittivity \eqref{eq:constitutive_relation} is  
\begin{subequations}\label{eq:eps_hat_circular}
\begin{equation}
    \hat{\varepsilon} = \mathrm{diag}(\varepsilon_\text{CRA},\ \varepsilon_\text{CRI},\ \varepsilon_\perp) \;,
\end{equation}
where
\begin{align}
    \varepsilon_\text{CRA} 
    = \varepsilon_\mathrm{bg} + \frac{i}{\omega} \sigma_\text{CRA} \;,
    \qquad 
    \varepsilon_\text{CRI} = \varepsilon_\mathrm{bg} + \frac{i}{\omega} \sigma_\text{CRI} \;,
    \qquad
    \varepsilon_\perp = \varepsilon_{\mathrm{bg}} + \frac{i}{\omega} \sigma_\perp \;.
\end{align}
\end{subequations}
Here, $\varepsilon_{\mathrm{bg}}$ is the background (ionic) permittivity, which is assumed to be homogeneous and isotropic. 
For the GaAs quantum wells that we consider, we take a value of $\varepsilon_\mathrm{bg} \approx 3.40^2 = 11.56$ at room temperature and THz frequencies~\cite{supplemental_adachi_gaas_1985}. At 300 mK temperatures, the refractive index is expected to decrease~\cite{supplemental_talghader_thermal_1995,supplemental_gehrsitz_refractive_2000}.
This has a small effect on the optimization and overall boost factor, and, additionally, $x$ may be modified to generate the same results as the room-temperature refractive indices, if desired.
For \SQWARE{}, the permittivity can be measured using ellipsometry or reflection/transmission spectroscopy~\cite{supplemental_chen_introduction_2022, supplemental_kriisa_cyclotron_2019}.  
Note that $\varepsilon_\text{CRA}$, $\varepsilon_\text{CRI}$, and $\varepsilon_\perp$ all depend upon the electron scattering time $\tau$ and the electron density $n_\text{bulk}$ (through $\sigma_0$), but only $\varepsilon_\text{CRA}$ and $\varepsilon_\text{CRI}$ depend upon the external magnetic field $|\Bvec_\mathrm{ext}|$.  
Therefore, the axion-induced electric field, which is parallel to the external field $\Evec_\mathrm{ax} \propto \hat{\varepsilon}^{-1} \Bvec_\mathrm{ext} \propto \Bvec_\mathrm{ext} / \varepsilon_\perp$ as in \erefs{eq:Ea}{eq:sigma_hat_linear}, only depends upon $\varepsilon_\perp$ and does not have a tunable response to $|\Bvec_\mathrm{ext}|$. 
For \SQWARE{}, the strength of the external magnetic field should be used to tune the resonant frequency, and the discussion here displays how this is not possible with a 3DEG.  
We next discuss how an anisotropic material allows for the desired response.  

The electromagnetic response of a quantum well (QW) can be modeled as a two-dimensional electron gas (2DEG)~\cite{supplemental_baer_two-dimensional_2015}.  
Electrons are restricted to move only within the QW, which can be approximated as a two-dimensional surface.  
Note that isotropy is broken by both the material and the external magnetic field.  
Rather than using a basis in which the $z$-axis is aligned with the external magnetic field, as we have done in the discussion of the 3DEG above, it is now convenient to use a basis in which the $z$-axis is aligned normal to the plane of the QW.  
In this basis, we can write the external magnetic field as 
\begin{align}
    \Bvec_\mathrm{ext} 
    & = 
    \underbrace{B_{\mathrm{ext},x} \, \evec_x + B_{\mathrm{ext},y} \, \evec_y}_{\ = \ \Bvec_{\mathrm{ext},\parallel}} + \underbrace{B_{\mathrm{ext},z} \, \evec_z}_{\ = \ \Bvec_{\mathrm{ext},\perp}} 
    = \underbrace{B_{\mathrm{ext},\text{CRA}} \, \evec_\text{CRA} + B_{\mathrm{ext},\text{CRI}} \, \evec_\text{CRI}}_{\ = \ \Bvec_{\mathrm{ext},\parallel}} + \underbrace{B_{\mathrm{ext},\perp} \, \evec_\perp}_{\ = \ \Bvec_{\mathrm{ext},\perp}} \;.
\end{align}
Note that $B_{\mathrm{ext},\text{CRA}} = \evec_\text{CRA}^\ast \cdot \Bvec_\mathrm{ext} = (B_{\mathrm{ext},x} - i B_{\mathrm{ext},y}) / \sqrt{2}$ and $B_{\mathrm{ext},\text{CRI}} = \evec_\text{CRI}^\ast \cdot \Bvec_\mathrm{ext} = (B_{\mathrm{ext},x} + i B_{\mathrm{ext},y}) / \sqrt{2}$ and $B_{\mathrm{ext},\perp} = \evec_\perp^\ast \cdot \Bvec_\mathrm{ext} = B_{\mathrm{ext},z}$.  
It follows that $|B_{\mathrm{ext},\text{CRA}}| = |B_{\mathrm{ext},\text{CRI}}| = \sqrt{(B_{\mathrm{ext},x})^2 + (B_{\mathrm{ext},y})^2} / \sqrt{2} = |\Bvec_{\mathrm{ext},\parallel}| / \sqrt{2}$.  
In this basis, the conductivity tensor is diagonal and takes the same form as in \eref{eq:sigma_hat_circular} with the replacements 
\begin{align}\label{eq:2DEG_replacement}
    \sigma_0 \to {n_{\mathrm{e}}} e^2 \tau / m^* d_\text{QW} 
    \qquad \text{and} \qquad 
    \omega_c \to e |B_{\mathrm{ext},\perp}| / m^* 
    \;,
\end{align}
where $n_{\mathrm{e}}$ is the surface carrier density (number per unit area).   
Notably, $\omega_c$ only depends upon $|B_{\mathrm{ext},\perp}|$, and the magnetic field components parallel to the surface of the quantum well $\Bvec_{\mathrm{ext},\parallel}$ do not contribute to the electronic motion, in contrast to the bulk 3DEG case, as the electrons are confined in the quantum well. 
For typical GaAs/AlGaAs QWs, only fields $\gg 10 \, \mathrm{T}$ could induce inter-well tunneling; for the field strengths that we consider, the Landau-level splitting $\omega_c \approx (0.063 \, \mathrm{eV}) ( |B_{\mathrm{ext},\perp}| / 36 \, \mathrm{T}) (m^* / 0.034 \, \mathrm{MeV})^{-1}$ is much smaller than QW potential depth. In the thin-film approximation, the electric permittivity takes the same form as in \eref{eq:eps_hat_circular} with the replacements noted above \cite{supplemental_li_vacuum_2018}.

The axion-induced electric field $\Evec_\mathrm{ax}$ is then calculated using \eref{eq:Ea}, which gives 
\begin{align}
    \Evec_\mathrm{ax}(\rvec,t) = - \gagg a_0 \biggl( \frac{B_{\mathrm{ext},\text{CRA}}}{\varepsilon_\text{CRA}(\zvec,m_\mathrm{a})} \, \evec_\text{CRA} + \frac{B_{\mathrm{ext},\text{CRI}}}{\varepsilon_\text{CRI}(\zvec,m_\mathrm{a})} \, \evec_\text{CRI} + \frac{B_{\mathrm{ext},\perp}}{\varepsilon_\perp(\zvec,m_\mathrm{a})} \, \evec_\perp \biggr) \, e^{-i m_\mathrm{a} t} 
    \;.
\end{align}
If we allow $\theta$ to denote the angle between the magnetic field and the surface normal of the 2DEG, then $|\Bvec_{\mathrm{ext},\parallel}| = |\Bvec_\mathrm{ext}| \sin\theta$ and $|\Bvec_{\mathrm{ext},\perp}| = |\Bvec_\mathrm{ext}| \cos\theta$.  
It follows that the dimensionless boost factor \eqref{eq:beta_def} in a QW is 
\begin{align}\label{eq:beta_QW}
    \beta = \frac{\|\Evec_\mathrm{ax}\|}{E_\mathrm{vac}} 
    = \sqrt{ \biggl( \frac{(\sin\theta)/\sqrt{2}}{|\varepsilon_\text{CRA}(\zvec,m_\mathrm{a})|} \biggr)^2 
    + \biggl( \frac{(\sin\theta)/\sqrt{2}}{|\varepsilon_\text{CRI}(\zvec,m_\mathrm{a})|} \biggr)^2 
    + \biggl( \frac{\cos\theta}{|\varepsilon_\perp(\zvec,m_\mathrm{a})|} \biggr)^2
    } 
    \qquad \text{(inside QW)} \;,
\end{align}
which quantifies the enhanced electric field strength inside the 2DEG.  
We discuss radiation out of the material in \sref{sub:radiation}.  

\begin{figure}[h]
    \centering
    \includegraphics[height=0.3\linewidth]{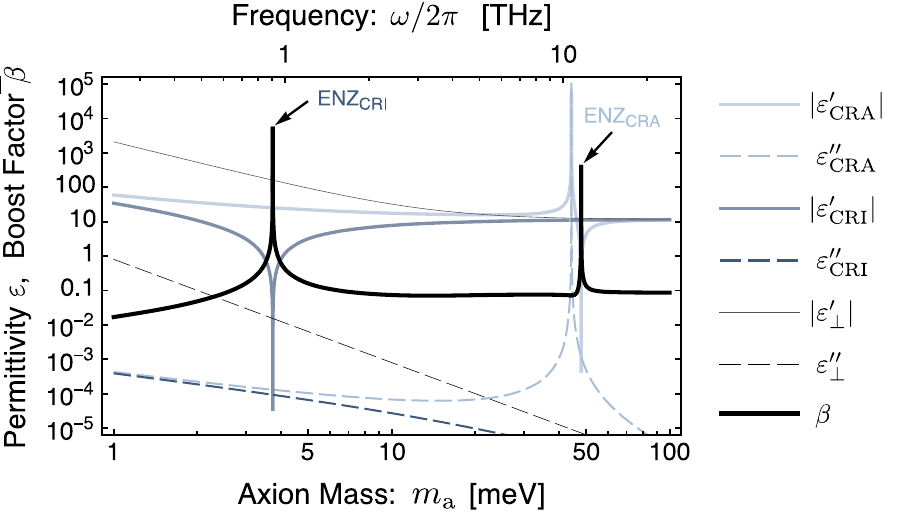}
    \caption{\label{fig:ENZ_point}
    The electric permittivity $\varepsilon$ and boost factor $\beta$ from \eref{eq:beta_QW} (inside the QW), as a function of the axion mass $m_\mathrm{a}$, showing the ENZ points.  In the circular polarization basis, the permittivity tensor is diagonal with entries $\varepsilon_\text{CRA}(\kvec,\omega)$, $\varepsilon_\text{CRI}(\kvec,\omega)$, and $\varepsilon_\perp(\kvec,\omega)$. We plot the magnitude of the real and imaginary parts, $\varepsilon = \varepsilon^\prime + i \varepsilon^\pprime$, evaluated at $\kvec = \zvec$ and $\omega = m_\mathrm{a}$.  We take the material parameters to be those of Config.\,2 and the magnetic field $|\Bvec_\mathrm{ext}| = 36 \, \mathrm{T}$ is oriented $\theta = 45\,\mathrm{deg}$ to the surface normal of the 2DEG. }
\end{figure}

In \fref{fig:ENZ_point} we show the three permittivities $\varepsilon$ and the boost factor $\beta$ as a function of the axion mass for a fiducial parameter set. 
The real part of $\varepsilon_\text{CRI}$ passes through zero at $m_\mathrm{a} \approx 4.23 \, \mathrm{meV}$, and the real part of $\varepsilon_\text{CRA}$ passes through zero at $m_\mathrm{a} \approx 43.9 \, \mathrm{meV}$.  
Close to these zero-crossing points are the epsilon-near-zero (ENZ) points where the signal boost is resonantly enhanced.  
The imaginary parts of the permittivity at the ENZ points are $\mathrm{Im}[\varepsilon_\text{CRI}] \approx 9 \times 10^{-5}$ and $\mathrm{Im}[\varepsilon_\text{CRA}] \approx 9 \times 10^{-4}$.  
Since the imaginary part of $\varepsilon_\text{CRI}$ is smaller at its ENZ point by a factor of $\approx 10$, a proportionally larger boost factor is generated.  
Consequently, the axion-induced signal is dominantly CRI-polarized, and we focus on this meV-mass region in the example configuration of \SQWARE{}.  
Although this figure shows signal boosts reaching values above $\beta = 10^3$, one should bear in mind that this is the boost factor for the electric field inside the 2DEG material.  
In the following sections, we discuss the propagation of electromagnetic radiation outside the material and a commensurate reduction in the boost factor to values $\beta \sim 10-100$ for example configurations.
\subsection{Multiple quantum well forming a plasmonic cavity}
\label{sub:MQWs}

Although an individual QW can achieve a large signal boost if the axion mass coincides with the ENZ point \eqref{eq:beta_QW}, the signal is suppressed by the tiny volume of a single QW.  
In the example configurations utilizing a GaAs/AlGaAs system, the typical QW thickness and width are $d_\text{QW} \sim 30 \, \mathrm{nm}$ and $W \sim 3 \, \mathrm{cm}$.  
The volume can be increased by layering quantum wells and barriers to form a stack of multiple quantum wells (MQWs), which forms an effective plasmonic cavity. 
The total thickness of the MQW is approximately $D= N_\mathrm{layer} (d_\text{QW} + d_\text{Barrier})$ where $N_\mathrm{layer}$ is the number of QW/barrier layers, $d_\text{QW}$ is the thickness of a QW layer, and $d_\text{Barrier}$ is the thickness of a barrier layer. 
The optimal choice for the cavity thickness (without a mirror) is approximately $D= \lambda_\mathrm{med} / 2$ where $\lambda_\mathrm{med} = 2 \pi / m_\mathrm{a} |\sqrt{\varepsilon_{\text{eff},\text{CRI}}}|$ is the wavelength of the CRI polarization mode at $\omega = m_\mathrm{a}$ in the medium.  
For smaller $D$ the signal is suppressed by the volume, and for larger $D$ the signal is suppressed by dissipative effects (see \sref{sub:radiation}). 

The effective permittivity of the MQW in the CRI polarization mode $\varepsilon_{\text{eff},\text{CRI}}(\kvec,\omega)$ may be calculated using effective medium theory (EMT) \cite{supplemental_haija_review_2011, supplemental_ferrari_hyperbolic_2015, supplemental_wu_frequency-tunable_2022}.  
This calculation amounts to an average over the layers of the MQW: 
\begin{equation}\label{eq:eps_eff}
    \varepsilon_{\text{eff},\text{CRI}} 
    = \frac{\varepsilon_{\text{CRI}} \, d_\text{QW} + \varepsilon_{\text{Barrier}} \, d_\text{Barrier}}
{d_\text{QW} + d_\text{Barrier}} \;.
\end{equation}
The CRI component of the electric permittivity in the QW layer $\varepsilon_{\text{CRI}}$ is calculated using \erefs{eq:eps_hat_circular}{eq:2DEG_replacement}.  
The isotropic electric permittivity in the barrier layer is $\varepsilon_{\text{Barrier}} = 3.27^2 = 10.6929$ for $\text{Al}_x\text{Ga}_{1-x}\text{As}$, where $x=0.24$, at room temperatures and THz frequencies~\cite{supplemental_adachi_gaas_1985}. 
The boost factor $\beta$ inside the MQW can be calculated using \eref{eq:beta_QW} with the replacement $\varepsilon_\text{CRI}(\zvec,m_\mathrm{a}) \to \varepsilon_{\text{eff},\text{CRI}}(\zvec,m_\mathrm{a})$, which gives 
\begin{align}\label{eq:beta_MQW}
    \beta 
    \approx \biggl| \frac{(\sin\theta)/\sqrt{2}}{\varepsilon_{\text{eff},\text{CRI}}(\zvec,m_\mathrm{a})} \biggr| 
    \qquad \text{(inside MQW / plasmonic cavity)} \;.
\end{align}
We neglect the $\evec_\text{CRA}$ and $\evec_\perp$ polarization modes, which is a good approximation near the ENZ point for the $\evec_\text{CRI}$ mode.  

Effective medium theory is an applicable description of the MQW provided that $d_\text{barrier},\ d_\text{QW} \ll \lambda_\text{med}$ such that each layer is thin relative to the electromagnetic wavelength in the medium. 
This condition is easily satisfied for the plasmonic cavities that we consider for an example \SQWARE{}.  
Note also that using a weighted average only applies to polarization modes that are normal to the stacking dimension (parallel to the quantum well surface), which is the case for the CRI (and CRA) component \cite{supplemental_popov_surface_2018}. 
We validate the EMT by comparison with direct numerical simulations of the MQW, and these results are reported in the Appendix. 
\subsection{Radiation propagating out of a plasmonic cavity}
\label{sub:radiation}
Although a strong signal boost can be achieved near an ENZ point inside of an MQW, in practice the enhanced electric field must propagate out of the plasmonic cavity in order to be measured.  
This propagation entails a reduction in the boost factor.  
Assuming that the plasmonic cavity has thickness $D$ and infinite transverse extent, and assuming that the region outside of the plasmonic cavity is in vacuum ($\hat{\varepsilon} = \hat{\mu}^{-1} = 1$), then the boost factor in the vacuum region is given by \cite{supplemental_millar_dielectric_2017} 
\begin{equation}\label{eq:beta_outside}
    \beta = \frac{\|\Evec_\mathrm{prop}\|}{E_\mathrm{vac}} 
    = \frac{\sin\theta}{\sqrt{2}} \, \biggl| \frac{1}{\varepsilon_{\text{eff},\text{CRI}}} - 1 \biggr| \, \biggl| 1 + \frac{i}{\sqrt{\varepsilon_{\text{eff},\text{CRI}}}} \cot\tfrac{\Delta}{2} \biggr|^{-1} 
    \qquad \text{(outside plasmonic cavity)} \;,
\end{equation}
where $\cot(x) = \cos(x) / \sin(x)$.  
The complex permittivity $\varepsilon_{\text{eff},\text{CRI}}(\zvec,m_\mathrm{a})$ is given by \eref{eq:eps_eff} and evaluated at $\omega = m_\mathrm{a}$.  
We also define the complex angular variable $\Delta = 2 \pi d / \lambda_\mathrm{med}$ and the complex effective wavelength in medium $\lambda_\mathrm{med} = 2 \pi / m_\mathrm{a} \sqrt{\varepsilon_{\text{eff},\text{CRI}}}$.  
Note that the second factor in \eref{eq:beta_outside} vanishes as $\varepsilon_{\text{eff},\text{CRI}} \to 1$, because we calculate $\beta$ using only $\Evec_\mathrm{prop}$ in \eref{eq:beta_def}, which is a good approximation when $\varepsilon_{\text{eff},\text{CRI}}$ is small.  
Note that the third factor tends to suppress $\beta$ relative to its value inside of the plasmonic cavity at the ENZ point \eqref{eq:beta_MQW}.  

The suppression is mitigated if $|\Delta| \lesssim \pi$ corresponding to $D\lesssim |\lambda_\mathrm{med}|/2$.  
Notice that the boost factor \eqref{eq:beta_outside} does not depend on the distance between the measurement point and the surface of the plasmonic cavity. 
This is a consequence of assuming that the plasmonic cavity has an infinite transverse extent.  
If the cavity has width $W$, where $W \sim \mathrm{cm}$ for the example \SQWARE{}, then we expect \eqref{eq:beta_outside} to be a good approximation if the electric field is measured at a distance of less than $\sim W$ from the surface of the plasmonic cavity. 
We validate this assumption and quantify its limitations using direct numerical simulation, which is reported in \sref{sec:COMSOL}.  
In practice, a lens is required to guide radiation toward photodetectors where it can be measured.  
However, if only a single photodetector is available, then even if the lens is able to efficiently redirect the radiation emitted from the front surface of the plasmonic cavity, the radiation from the back surface is lost.  
It would be possible to mitigate the situation by attaching a conducting sheet to the back side of the plasmonic cavity.  
The conductor would act as a mirror, redirecting more radiation toward the front of the cavity and toward the photodetector.  
In this configuration, the boost factor is further enhanced to 
\begin{equation}\label{eq:beta_mirror}
    \beta 
    = \frac{\sin\theta}{\sqrt{2}} \biggl| \frac{1}{\varepsilon_{\text{eff},\text{CRI}}} - 1 - \frac{1}{\varepsilon_{\text{eff},\text{CRI}}} \sec\Delta \biggr| \biggl| 1 - \frac{i}{\sqrt{\varepsilon_{\text{eff},\text{CRI}}}} \tan\Delta  \biggr|^{-1} 
    \qquad \text{(outside plasmonic cavity with mirror)} 
    \;.
\end{equation}
For more complex multi-layer architectures, further enhancement can be pursued through numerical optimization or machine learning techniques~\cite{supplemental_jung_simulation_2022, supplemental_mcdonald_scanning_2022}. 
\subsection{Scanning axion masses with variable magnetic field orientation}
\label{sub:scanning}
For \SQWARE{}, the orientation of the magnetic field relative to the plasmonic cavity impacts the resonant frequency at the ENZ point. 
If the axion mass coincides with this resonant frequency, the resultant electromagnetic radiation is enhanced by a boost factor \eqref{eq:beta_outside} that can be much larger than $1$.  
In order for the experiment to search for axion DM across a range of masses, the plasmonic cavity will be attached to a mount that can be rotated, thereby allowing the relative angle of the plasmonic cavity and the external magnetic field to be varied freely.  
See \sref{sub:magnet} for a discussion of the experimental apparatus. 

\begin{figure}[H]
    \centering
    \includegraphics[height=0.33\linewidth]{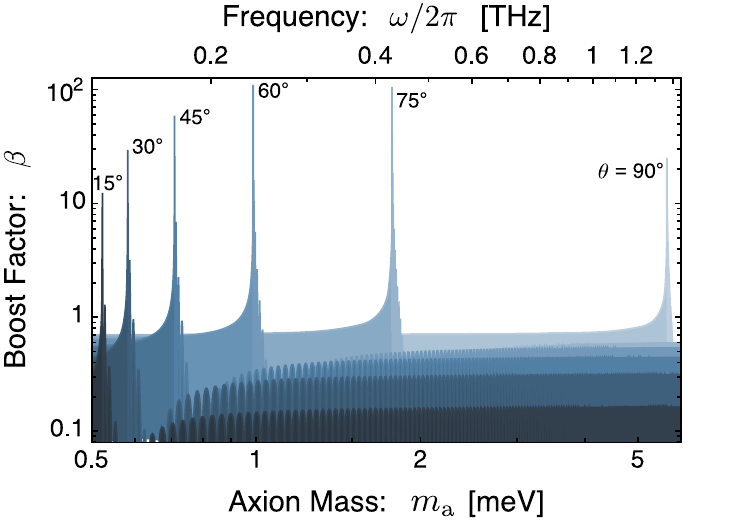}\hspace{0.4cm}\includegraphics[height=0.36\linewidth]{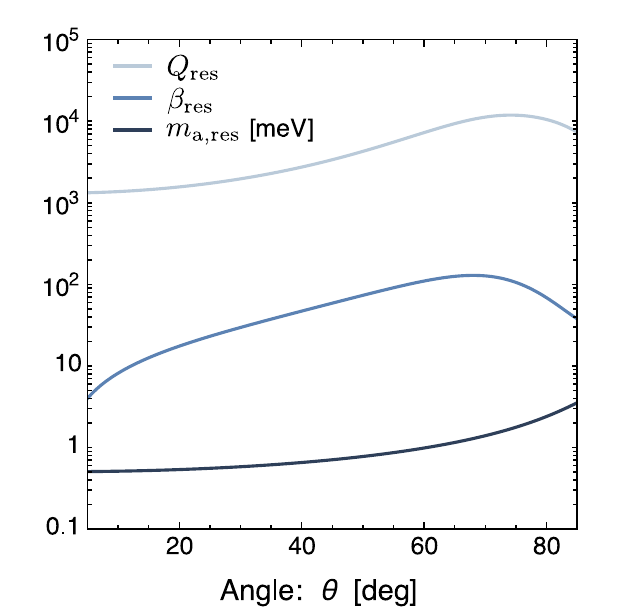}
    \caption{\label{fig:vary_theta}
    Using Config.\,2 parameters, as the angle $\theta$ between the external magnetic field $\Bvec_\mathrm{ext}$ and the surface normal of the plasmonic cavity is varied, the ENZ point, at which the CRI polarization mode is resonantly enhanced, scans across a range of axion masses. The \textit{left} figure plots boost factor $\beta$ from \eref{eq:beta_outside} (outside the plasmonic cavity) as a function of axion mass $m_\mathrm{a}$ for several values of $\theta$.  The mass at which $\beta$ spikes is at the resonantly-enhanced ENZ point for each $\theta$. The \textit{right} figure plots the value of the axion mass $m_{\mathrm{a},\text{res}}$ at resonance, the FWHM $m_{\mathrm{a},\text{res}}/\Delta m_{\mathrm{a},\text{res}}$ of the resonance, and the value of the boost factor $\beta_\text{res}$ at resonance, for several values of $\theta$ (At $\theta=0$, the boost factor from the CRI mode is 0). For each frequency or axion mass, a unique angle is optimal for maximum signal boost. The overall sharpest resonance (maximum $Q$ and minimum FWHM) determines the maximum scan rate.
    }
\end{figure}

In \fref{fig:vary_theta} we illustrate how the orientation of the magnetic field can be used to scan across axion masses.  
When the angle $\theta$ between the external magnetic field $\Bvec_\mathrm{ext}$ and the surface normal of the plasmonic cavity changes, the ENZ point in the CRI polarization mode shifts to a different resonant frequency, which probes a different axion mass.  
This behavior is illustrated in the left panel, which shows $\beta$, calculated from \eref{eq:beta_outside}, as a function of $m_\mathrm{a}$ for several different values of $\theta$.  
At each value of $\theta$, the spike in $\beta$ corresponds to the value of $m_\mathrm{a}$ that is resonantly enhanced due to the ENZ point in the CRI polarization mode.  
In the right panel we show the axion mass $m_{a,\text{res}}$ that is resonantly enhanced at each value of $\theta$, the corresponding value of the boost factor $\beta_\text{res}$ on resonance, and the quality factor or inverse of the normalized width of the resonance $Q=m_{a,\text{res}}/\Delta m_{a,\text{res}}$.  
These correspond to the local maximum of $\beta(m_\mathrm{a})$ given by \eref{eq:beta_outside} and its full width at half maximum (FWHM).  
As the angle varies, the resonance scans across axion masses from $m_\mathrm{a} \approx 0.5$ to $5$\,meV.  

The \SQWARE{} example configurations can scan across axion masses by varying the angle $\theta$ of the MQW within the magnetic field in discrete steps.  
For example, in Config.\,2, at $\theta =  72.437$\,deg, the boost factor is resonantly enhanced in a range of axion masses around $m_\mathrm{a} \approx 1.55535 \pm 0.00015$\,meV.  
The experiment will remain in this position for a time $t_\text{obs}=34\, \text{min}$ while data is collected.  
If no photons are detected, a limit can be placed on the size of the axion-photon coupling in this mass window.  
The angle is then incremented to $\theta = 72.439$\,deg, shifting the resonance by approximately one FWHM to $m_\mathrm{a} \approx 1.55550$\,meV.  
In this way, \SQWARE{} in general can search for axion DM across a continuously connected range of axion masses.  
The total time required to cover a window of axion masses from $m_{a,\text{min}}$ to $m_{a,\text{max}}$ is roughly $t_\text{tot} = t_\text{obs} \times (m_{a,\text{max}} - m_{a,\text{min}}) / \Delta m_{a,\text{res}}$.  

To draw the projected sensitivity curves, we select a range of masses $m_{a,\text{min}}$ to $m_{a,\text{max}}$, and we determine the minimum of $\Delta m_{a,\text{res}}$ across this range to use in this expression.  
Note that for all the parameters that we consider, the bandwidth of the photodetector is wider than the FWHM of the resonance; see the discussion in \sref{sub:photodetector}. 
\begin{figure}[H]
    \centering
      \includegraphics[height=0.33\linewidth]{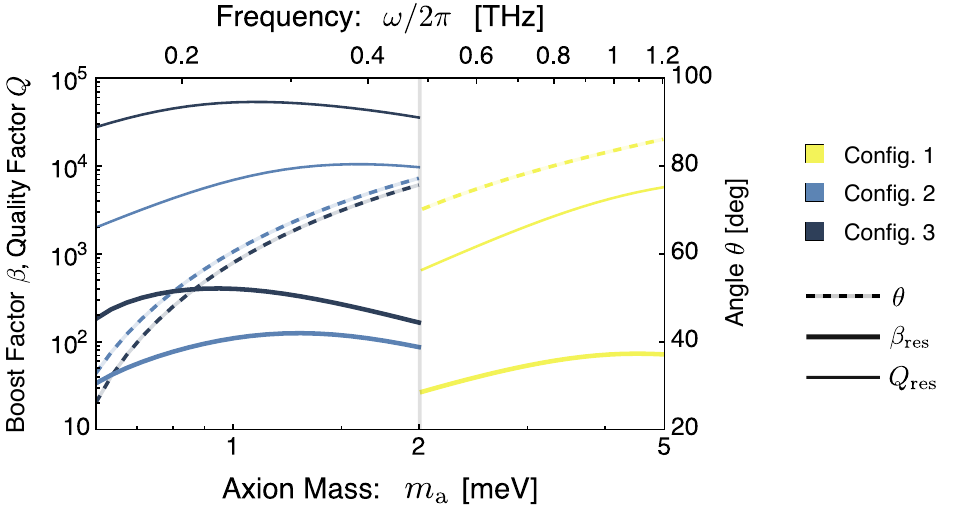}
    \caption{\label{fig:all configs theta, beta, q}
    For each configurations (Config.\,1 in yellow, Config.\,2. in blue, and Config.\,3 in navy) the optimal angle $\theta$ (dashed line) is swept across axion masses, maximizing the boost factor $\beta_\text{res}$ (thick solid line) and corresponding quality factor $Q_\text{res}$ (thin solid line) on resonance.
    }
\end{figure}

\section{COMSOL simulations of MQW electromagnetic response}
\label{sec:COMSOL}

\subsection{Implementing axion in COMSOL}
\label{sec:implementing}
To quantitatively assess the response of the MQW structure to the axion and validate analytical sensitivity calculations, we performed simulations using COMSOL Multiphysics®, a finite element method (FEM) solver and simulation package. The calculations validate the analytical analysis using effective medium theory (EMT), probe the impact of electron density, magnetic field, and spatial inhomogeneities, and inform design optimizations for maximizing the axion-photon conversion signal. The simulations are performed in the frequency domain, where all relevant fields are oscillating at the same frequency $\omega$ in time. All simulations are evaluated at a fixed frequency $\omega/2\pi =1$\,meV and in the CRI polarization basis.
The axion-induced electric field $\Evec_\mathrm{med}(\rvec, t)$ is modeled and inputted into COMSOL as a constant background electric field, with frequency $\omega$, in each material domain within the simulation, given by \eref{eq:axion-induced electric field}, where $\varepsilon$ is a function of the material geometry and is kept as a scalar for simplicity. For example, across the MQW structure, $\varepsilon = \varepsilon_\text{CRI}$ and across the vacuum $\varepsilon=1$. An alternative approach is to inject a background external current density $\Jvec_\mathrm{ax} = - \gagg a_0 m_\mathrm{a} \Bvec_\mathrm{ext} \, e^{-i m_\mathrm{a} t} $ instead; both yield the same solution~\cite{supplemental_Gardikiotis2020_presentation}. COMSOL then solves for the scattered electric field $\Evec(\rvec)$ by solving the following equations (in the frequency domain)
\begin{subequations}
    \begin{equation}
        \nabla \times \frac{1}{\mu(\rvec)}(\nabla \times \Evec(\rvec))-k_0^2\varepsilon(\rvec)\Evec(\rvec) = 0
    \end{equation}
    where the permeability is set to $\mu(\rvec)=1$ and  $k_0$ is the wavenumber in a vacuum. The Ansatz are the homogeneous solutions of the Maxwell equations unchanged by axions as shown in \eref{eq:Maxwell_axion}.
    \begin{equation}
    \Evec(\rvec)= \tilde\Evec(\rvec)e^{-i\kvec \cdot\rvec},
    \end{equation}

\end{subequations}

Scattering boundaries are set far away ($\sim10 \, \lambda_\text{vac}$) from the MQW structure and satisfy the following conditions
\begin{subequations}
\begin{align}
    \nvec \times (\nabla \times &\Evec) - ik \nvec\times(\Evec \times \nvec) = 0 \\
    &\Evec_\text{total} = \Evec_\text{med} + \Evec
\end{align}
\end{subequations}
allowing for plane waves of first order from inside the simulation to scatter without reflection, also denoted as $E_\text{prop}$. Additionally, the magnetic field across the surface of the MQW is constant and drops slowly to zero outside the MQW in the vacuum, in order to prevent scattered fields generated at the boundaries.
\begin{figure}[H]
\centering
\includegraphics[height=0.25\linewidth]{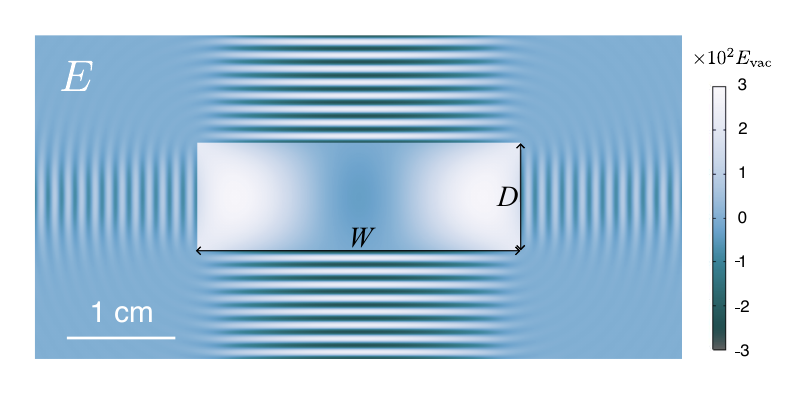}
\hspace{0.1cm}
\includegraphics[height=0.25\linewidth]{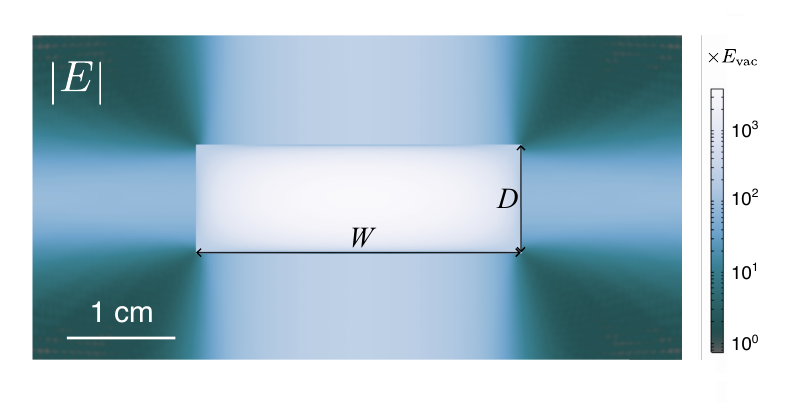}
\caption{COMSOL simulations for the example \SQWARE{} geometry (shown in 2D) allowing analysis of realistic experimental conditions. The plasmonic cavity with width $W$ is created from MQWs stacked over thickness $D$. The boost factor is averaged over a length $W$ displaced $1$\,cm above and below the cavity. The \textit{left} figure plots the electric field (out of the plane) and the \textit{right} figure plots the magnitude.}
\label{diagram of sim}
\end{figure}

Unless denoted otherwise, the simulated MQW structures assume Config.\,2 parameters with the modification $d_\text{Barrier}=3 d_\text{QW}$ (instead of $d_\text{Barrier}=5d_\text{QW}$), which only slightly shifts the resonance magnetic field and boost factor.
The magnetic field $\Bvec_\mathrm{ext}$ is in a fixed orientation at an angle $\theta=52$\,deg, relative to the MQW surface normal. For some simulations, the angle is tuned in increments on the order of $0.001$\,deg to achieve the maximum boost factor, as certain simulated experimental conditions (such as the finite size of the MQW, finite layer thickness, and spatial inhomogeneities) additionally shift the resonance magnetic field in comparison to ideal analytical values, and the maximum achievable boost factor remains relatively unchanged, as the analysis that follows will demonstrate. Interestingly, the cyclotron resonance frequency itself may also deviate, due to hybridization with the low-frequency plasmonic resonance within the material, and again lead to slightly shifted resonance magnetic fields and signal boosts~\cite{supplemental_kriisa_cyclotron_2019}. In any case, \textit{in~situ} tunability (with a tilting probe or by directly modifying the strength of the magnetic field) is needed for accurate calibration of the axion mass if a signal is detected.
The boost factor $\beta$ in simulation is defined as the ratio of the amplitude of surface-emitted propagating or scattered radiation to the amplitude of the axion-induced field in vacuum, $E_\text{vac}$, averaged over a surface patch displaced $1$\,cm normal to and coextensive with the MQW surface; since the MQW is simulated in the CRI basis, the factor of $\sin\theta/\sqrt{2}$ from \eref{eq:beta_outside} is added in post-processing.
\subsection{Beyond effective medium theory (EMT)}
\label{subsec:emtvalid}
\begin{figure}[H]
\centering
\includegraphics[width=0.95\linewidth]{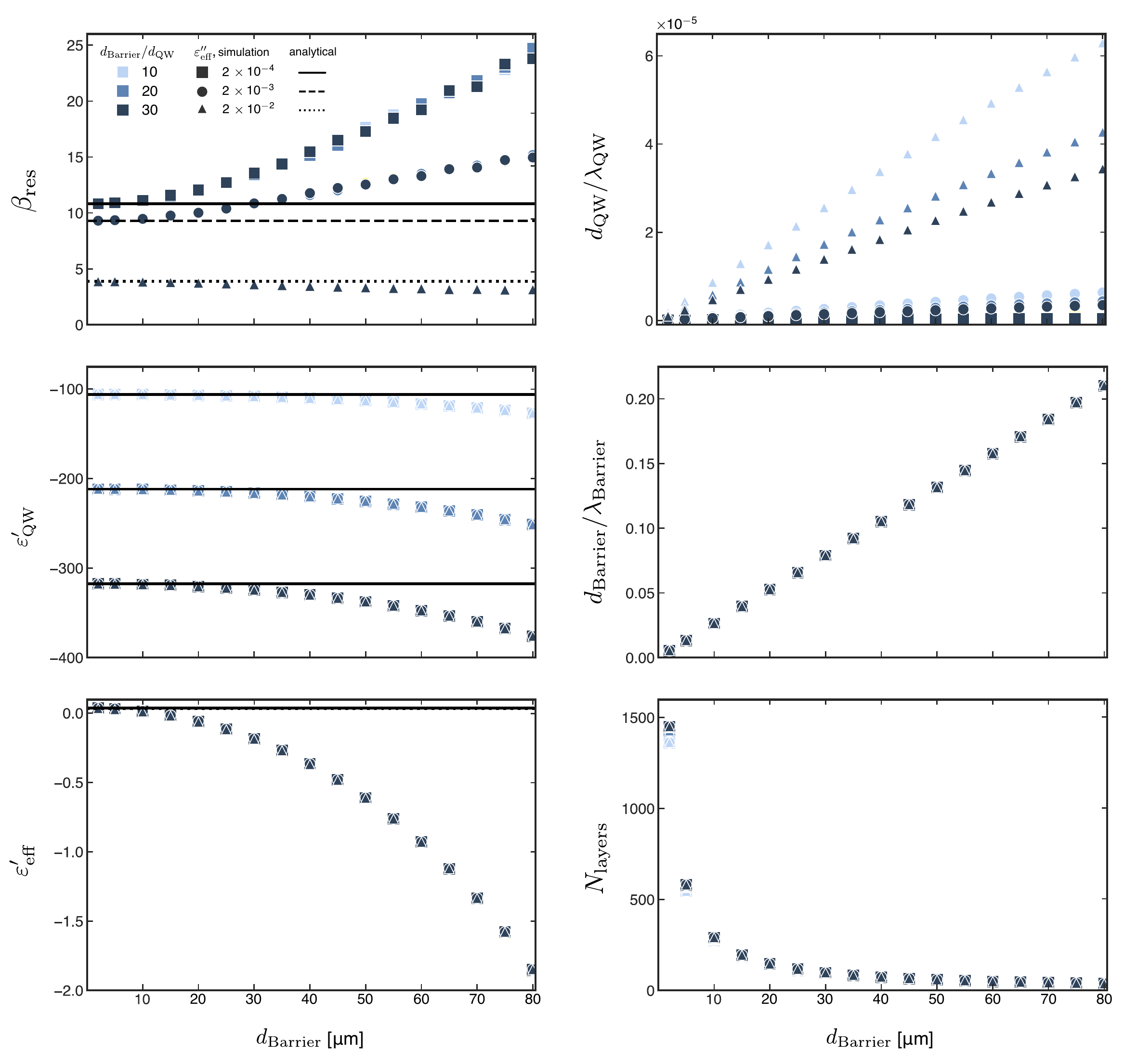}
\caption{Simulations of an example MQW structure with fixed thickness $D= 0.3$\,cm at $1$\,meV, oriented at an angle $\theta=42$\,deg, while varying the real part of the permittivity in the quantum well $\varepsilon'_\text{QW}$ and the layer structure. The \textit{top left} figure shows the maximum boost factor on resonance, $\beta_{\text{res}}$, as a function of barrier layer thickness for different ratios $d_\text{Barrier}/d_\text{QW}$ (with light as 10, blue as 20, and dark blue as 30) and imaginary part of the effective permittivity $\varepsilon''_\text{eff}$ or losses (with dots as $2\times 10^{-4}$, crosses as $2\times 10^{-3}$, squares as $2\times 10^{-2}$). The expected analytical values of the boost factor for the different  $\varepsilon''$, or losses, are also plotted in horizontal black lines (with solid as $2\times 10^{-4}$, dashed as $2\times 10^{-3}$, and dotted as $2\times 10^{-2}$. Varying the ratio did not significantly affect the maximum boost factor. As the losses increase, the boost factor decreases in both analytical calculations and simulations, as expected. As the barrier thickness increases, the simulated maximum boost factor diverges from the analytical expression. 
As the barrier thickness increases, the maximum boost factor in simulations occurs at more negative values of $\varepsilon'_\text{QW}$, shown in the \textit{middle left} figure, and therefore $\varepsilon'_\text{eff}$, shown in the \textit{bottom left}. The breakdown from EMT is further demonstrated in \textit{top right} and \textit{middle right} figures, where the layer thicknesses are increased relative to the wavelength in the medium, either the quantum well or the barrier, respectively. The barrier thicknesses have a more significant effect on EMT breakdown, as $d_\text{Barrier}/\lambda_\text{Barrier}$ is much greater than $d_\text{Barrier}/\lambda_\text{Barrier}$ in all simulations. The total number of layers $N_\text{layers}$ is also plotted in the \textit{bottom right}.}
\label{emtenhance}
\end{figure}

Even outside the strict effective medium theory (EMT) limit, significant boost factors are possible in simulations at shifted resonance conditions compared to EMT, further illustrated in \fref{emtenhance}. For a hypothetical material with thicker/fewer layers and similar losses, EMT breaks down, and the boost factor reaches much higher than the analytically predicted maximum for a homogeneous medium, and the resonance is optimized at significantly more negative permittivities, which is challenging for 2DEGs while keeping losses minimal, as seen in Fig.~3 in the main text. In addition, thicker GaAs layers will behave as a bulk 3DEG rather than 2DEG, so the magnetic field can not be used for tuning anymore, as outlined in~\sref{sub:cyclotron}~\cite{supplemental_ando_electronic_1982, supplemental_gilbertson_dimensional_2009}. 
\subsection{Finite area of plasmonic cavity}
\label{subsec:finitearea}
\begin{figure}[H]
\centering
\includegraphics[height=0.32\linewidth]{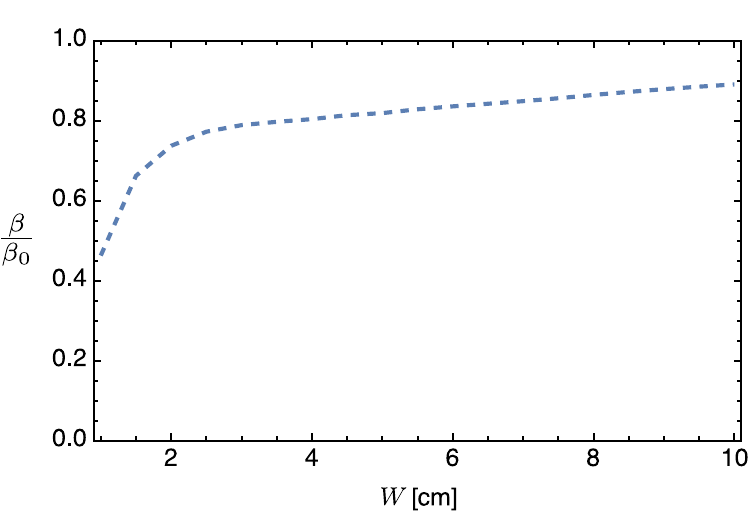}
\caption{Relative boost factor versus MQW structure surface width $W$ for fixed thickness $D=1$\,cm. Boost factor approaches the maximum 1D/infinite area value at larger aspect ratios, demonstrating finite-size suppression for small samples.}
\label{fig:var4}
\end{figure}

The analytical boost factor from \eref{eq:beta_outside} assumes an MQW with infinite surface and is solved as a 1D equation, giving $\beta_0$. In greater dimensions, where the material now has a finite surface area, the surface radiation is less collimated, which reduces the effective signal boost. In a 2D COMSOL simulation, the axion-induced electric field is averaged over a surface with varying width $W$, capturing the effect of finite MQW size on the collimation and intensity of surface-emitted radiation. An MQW surface width of $W=3$\,cm for $D=1$\,cm, in Config.\,2, is sufficient to maintain at least 50\% of the boost factor calculated analytically, as illustrated in \fref{fig:var4}. For the 2D simulations in the following sections, the relative boost factor $\beta/\beta_0$ will take $\beta_0$ as the boost factor for aspect ratio $W/D=1.5$.  

\subsection{Axion inhomogeneities}
\label{sub:axionhom}
\begin{figure}[H]
\centering
\includegraphics[height=0.32\linewidth]{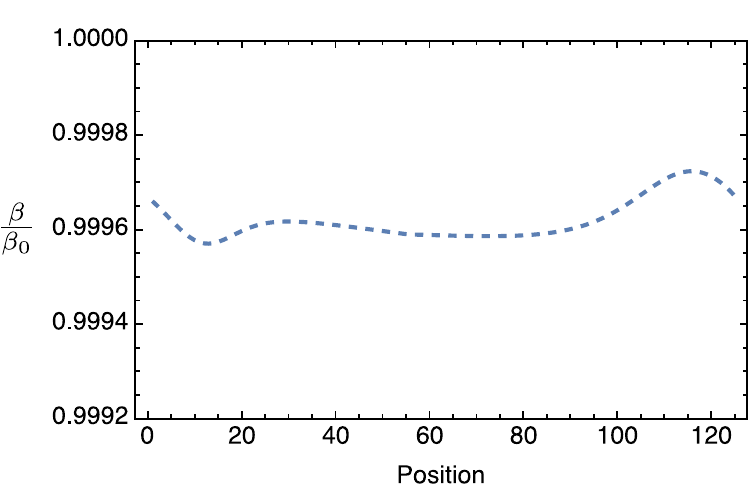}
\caption{Finite axion spatial gradient (where the axion field is modeled here as a random function of position) only slightly modifies the boost factor by less than a fraction of a percent.}
\label{fig:coh}
\end{figure}

The simplified axion-modified Maxwell equations (\eref{eq:Maxwell_axion} assume a homogeneous axion field with fixed amplitude. DM axions, however, vary in space and time. This means the axion field will exhibit a non-zero gradient or finite coherence length and time~\cite{supplemental_adams_axion_2023}. The coherence time on the order of $\mu$s for meV axions, much longer than the detector response, which is on the order of ns~\cite{supplemental_kono_picosecond_1999}, and the coherence length, calculated from the axion deBroglie wavelength, is on the order of 10s of cm for meV axions, much larger than the detector, which is on the order of a few cm. The spatial and time-varying axion field is modeled using Rayleigh distributed random variables~\cite{supplemental_amaral_vector_2024}. 125 simulations are performed by ``sweeping'' the detector across the spatial axion field at a particular moment in time. \fref{fig:coh} normalizes the average axion field across the MQW for individual simulations to demonstrate only the effects of a finite spatial gradient.

\subsection{B-field inhomogeneities}
\label{sub:Binhom}
\begin{figure}[H]
    \centering
    \includegraphics[height=0.32\linewidth]{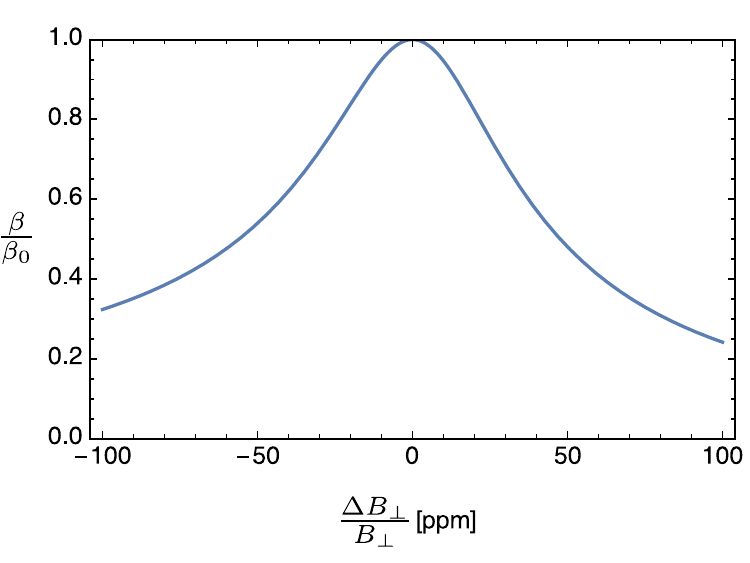}
    \caption{\label{fig:boostvbfield}
     Boost factor as a function of the variation in the magnetic field component normal to the surface of the MQW, $B_\perp$, centered at the resonant magnetic field. An inhomogeneous magnetic field that shifts the field from the resonance point will result in lower boost factors at a fixed frequency.}
\end{figure}

The boost factor as a function of the magnetic field at a specific frequency exhibits a sharp resonance, as shown in Figure~\ref{fig:boostvbfield}. Therefore, in order to maintain resonance over the entire MQW, a high-homogeneity magnetic field must be employed, which is easily possible with the $1$ ppm magnets available in experiment~\cite{supplemental_brandt_national_2001}. For magnets with worse homogeneity, the expected boost factor is reduced as the resonance and collimation are reduced across the emitting surface. 

\subsection{Electron density nonuniformity}
\label{sub:enonun}
\begin{figure}[H]
\centering
\includegraphics[height=0.32\linewidth]{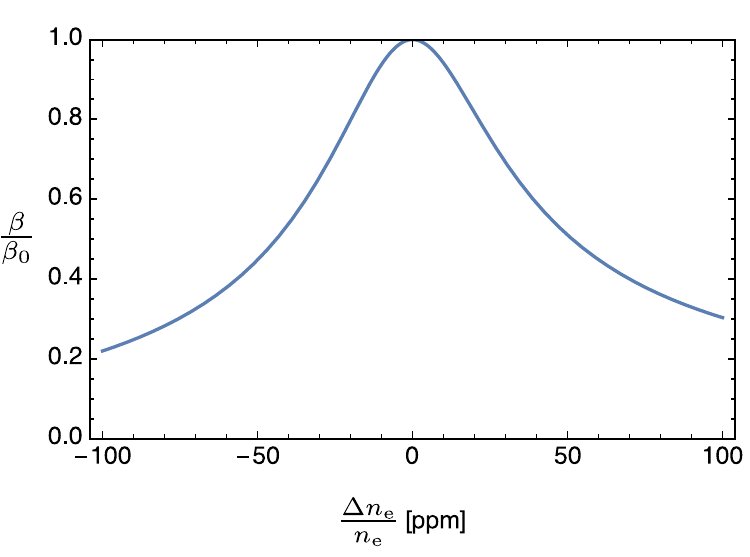}
\caption{Boost factor as a function of the electron density variation, centered at the resonant electron density. An inhomogeneous electron density that shifts the density from the resonance point will result in lower boost factors at a fixed frequency.
}
\label{fig:var6}
\end{figure}

\begin{figure}[H]
\centering
\includegraphics[height=0.3\linewidth]{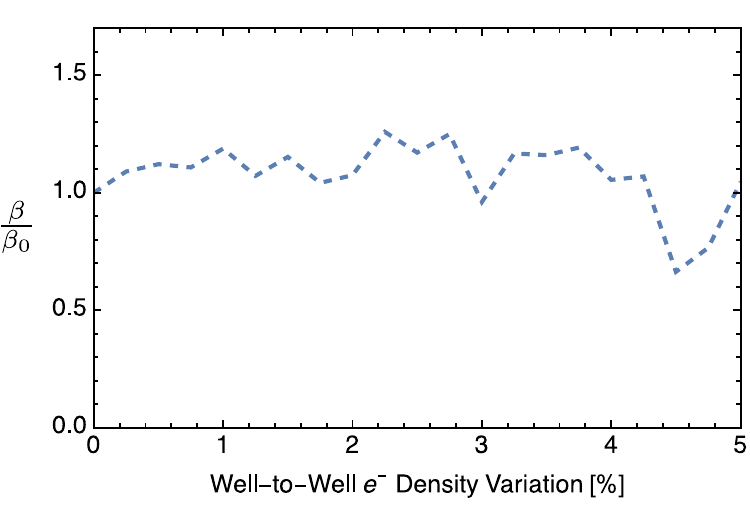}\label{fig:var2}\hspace{0.5cm}
\includegraphics[height=0.3\linewidth]{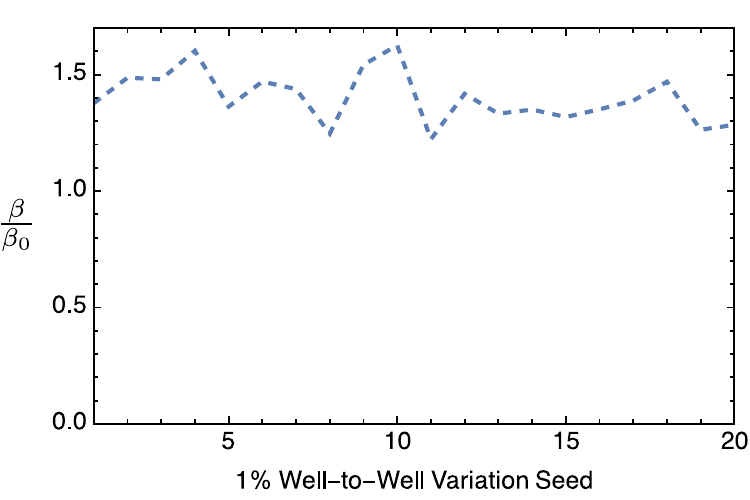}\label{fig:var7}
\caption{The \textit{left} figure plots the maximum relative boost factor from the versus well-to-well electron density variance. As variance increases, the maximum boost factor varies, but is not significantly degraded til 5\% well-to-well variance. The \textit{right} figure plots the maximum relative boost factor versus seed with fixed well-to-well electron density variance of 1\%. Out of 20 random seeds simulated, none show significant degradation of the maximum boost factor.}
\end{figure}

\begin{figure}[H]
\centering
\includegraphics[height=0.3\linewidth]{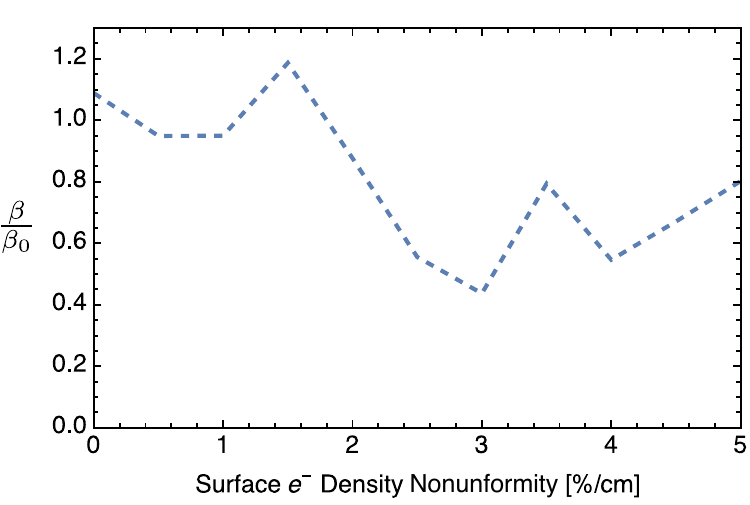}\label{fig:var3}
\hspace{0.5cm}
\includegraphics[height=0.3\linewidth]{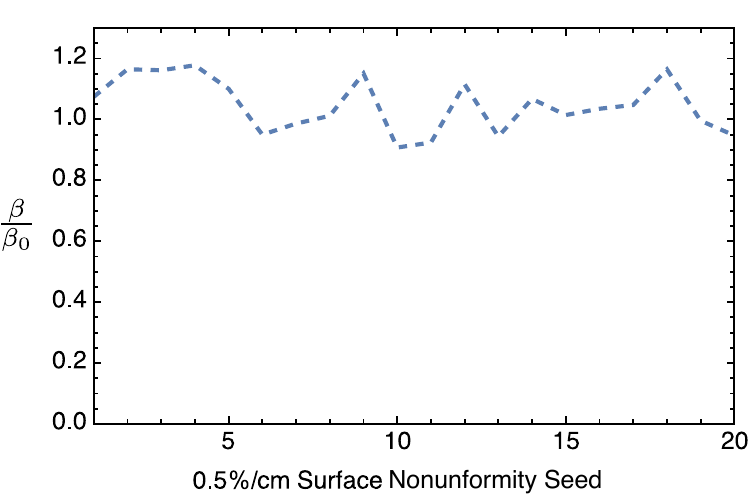}\label{fig:var8}
\caption{The \textit{left} figure plots the maximum relative boost factor versus electron density variation across the MQW surface in 2D simulation for a fixed well-to-well variation of 1\%. Larger inhomogeneity suppresses the boost factor by reducing phase coherence. The \textit{right} figure plots the maximum relative boost factor versus seed for a fixed average surface gradient of 0.5\%/cm and a fixed well-to-well variation of 1\%. Out of 20 random seeds simulated, none show significant degradation of the maximum boost factor. }
\end{figure}

The boost factor as a function of doping or electron density exhibits a sharp resonance, as shown in \fref{fig:var6}. Therefore, in order to maintain resonance over the entire MQW, the doping density must exhibit consistency. Simulations with electron well-to-well density variation up to 5\% and surface density variation up to 5\%/cm (equivalent to $5\times10^4$ ppm/cm) are conducted to explore sensitivity. Variations in the electron density $n_\mathrm{e}$ throughout each QW and across different QWs may result in reduced coherence of the ENZ point and smaller effective volumes of plasmonic material resonant at one common frequency. The boost factor is computed as a function of the density nonuniformity, using both 1D and 2D simulations. 
In the 1D simulation, the density varies well-to-well randomly using a discrete normal distribution. Figure~\ref{fig:var2} shows that random well-to-well electron density variation gradually reduces the maximum boost factor as the standard deviation of the variation increases. Due to randomly generated anisotropy, the boost factor may reach values above the maximum analytically calculated boost; plotted is the boost factor from the ``best'' side of the magnetoplasmonic cavity, which can be determined again through ellipsometry or spectroscopy~\cite{supplemental_chen_introduction_2022, supplemental_kriisa_cyclotron_2019}. A standard deviation of 1\% in the experiment is expected~\cite{supplemental_saito_growth_1987,supplemental_szerling_mid-infrared_2009}, which maintains a high boost factor in simulations conducted for 20 random seeds, shown in Figure~\ref{fig:var7}.
In the 2D simulations, a 1\% well-to-well average density variation is selected, but now with the addition of a surface density nonuniformity in each well. The surface variation of each well is approximated as a linear gradient with a random direction, and the magnitude of the gradient in each well is generated from a normal distribution with a standard deviation equivalent to the average gradient. Figure~\ref{fig:var3} plots the maximum boost, again from the ``best'' side, as a function of the average electron density gradient across the MQW surface.
In experiment, the deliberate randomization of the gradient from well-to-well can be accomplished during fabrication with substrate rotation and manipulation to randomize the orientation of the atomic beam whilst growing the sample, which is also used to achieve higher uniformity in each well as well~\cite{supplemental_saito_growth_1987,supplemental_gardner_modified_2016}. Simulating multiple random seeds with the mean gradient of 0.5\%/cm, which is approximately the expected value in the experiment, does not significantly degrade the boost factor, as shown in Figure~\ref{fig:var8}, due to effective medium theory incorporated with deliberate randomization. A single quantum well sample achieved a linear gradient of approximately 0.5\%/cm utilizing an angled beam and \textit{in~situ} rotation during growth~\cite{supplemental_chung_spatial_2019}. We can expect even better uniformity from the higher mobility samples considered for the example \SQWARE{}~\cite{supplemental_manfra_molecular_2014, supplemental_chung_ultra-high-quality_2021}. 

\textit{Summary}---Across all cases, the analytical boost factor expressions remain accurate for experimentally realistic layer thicknesses, finite surface sizes, and inhomogeneous scenarios. Under realistic experimental conditions, the maximum achievable boost factor remains above 50\% of the ideal value, confirming the robustness of this particular \SQWARE{} design. These simulations justify the use of analytical sensitivity projections and guide tolerances for material design and fabrication, as well as magnetic fields in future prototypes. Prior to fabrication and experimentation, some parameters such as the magnetic field profile and the expected area of the MQW surface can be used directly to estimate the adjusted boost factor; however, others, like the random electron density variation, require Monte Carlo or other robust statistical testing methods in order to accurately place limits on the systematic uncertainty. After fabrication, during experimentation, the boost factor and thus resonance quality factor may be exactly determined (see \sref{calibration}).
\section{Experimental design and operation}
\label{sec:expdesign}
In this section, we provide additional details regarding the design and operation of the \SQWARE{} detection strategy. 
We discuss each of the four core subsystems in turn:  the MQW (multiple quantum well) structure, the high-field magnet, the focusing lens, and the photodetector.  
We also discuss an optional conducting mirror that may be used to guide THz radiation onto the photodetector.  
We further discuss aspects of the experimental operation, such as calibration and signal acquisition. 
\subsection{MQW structure}
\label{sub:MQW_structure}
Multiple Quantum Wells (MQWs) are a mature platform in condensed matter physics, with established, high-quality fabrication techniques including Molecular Beam Epitaxy (MBE). Prototypes can be rapidly realized using widely available infrastructure, making the experiment highly accessible to a broad range of laboratories. The example configuration in this paper employs GaAs quantum wells (QWs) with Si monolayer doping, separated by AlGaAs barriers. 
Design parameters include the lateral dimensions $W \times W$, the total MQW structure thickness $D$, the individual QW thickness $d_\text{QW}$, the barrier thickness $d_\text{Barrier}$, the surface electron density $n_\mathrm{e}$, the scattering rate $\tau$, the electron mobility $\mu$, and the operating temperature $T$.  
For axion detection in \SQWARE{}, the signal power scales linearly with the MQW surface area ($A = W^2$), motivating the use of the largest practical wafers. 
Current MBE techniques provide for the fabrication of high-quality wafers up to $W \sim 5$\,cm~\cite{supplemental_chung_ultra-high-quality_2021}. 
The thickness $D$ should be approximately equal to half a wavelength in the plasmonic cavity, which is around $1$ to $10$\,mm at THz frequencies for the example \SQWARE{}. 
This would be larger than what is typically grown, where a recent state-of-the-art MQW with $166$ wells achieved $D\sim 0.01$\,mm, but quantum cascade lasers with $D\sim 0.025$\,mm demonstrate a proof of principle~\cite{supplemental_hale_multi-mode_2025,supplemental_li_multi-watt_2017, supplemental_sharma_study_2025}. Although these MQWs are thinner than those proposed for SQWARE, there is no physical limit on the number of quantum wells. The growth of thicker cavities will require significant investment and modification of standard MBE growth protocol, in order to maintain high quality and structural integrity when growing a single thick cavity. An alternative, more feasible, approach would involve growing several thinner MQW cavities on separate substrates in parallel and adhering them in series using a direct wafer bond process~\cite{supplemental_brandstetter_thz_2012}. Less costly prototypes may also be engineered in labs with limited magnetic field resources and unique growth constraints. For example, larger axion masses can be probed using smaller magnetic fields and thinner cavities resonating at shorter wavelengths. Thinner cavities will also be ideal to minimize radiation loss if using lower quality MBE protocols or materials with lower mobilities. Moreover, a given sample can be tested in upgraded magnetic fields as newer technologies become available. These smaller prototypes can make preliminary constraints in the axion parameter space, in addition to providing proof-of-concept before investing in larger structures.
The aspect ratio (lateral size $W$ vs. thickness $D$) further influences the collimation and collection efficiency of the emitted radiation, which requires $D \ll W$.
We quantify this effect in \sref{sec:COMSOL} using numerical simulation.
For the example \SQWARE{}, the ratio of the QW thickness $d_\text{QW}$ (fixed at $30$\,nm) and barrier thickness $d_\text{Barrier}$ is chosen to maximize the boost factor $\beta$.  
The expected fabrication time for an MQW, such as the one described here, would be weeks to months, based on a $1 \, \mu$m/hr growth rate \cite{supplemental_li_mbe_2015}. 
The electron density of the quantum well $n_\mathrm{e}$ is determined by the distance between the silicon doping layer and well layer. Values of $1 -3\times 10^{11}$\,cm$^{-2}$ are common for lightly-doped GaAs QWs~\cite{supplemental_li_vacuum_2018}. We selected the higher end of the standard, so we can use thicker and fewer barrier layers to achieve the same ENZ point at a particular frequency. Higher densities are possible by bringing the doping layer closer to the quantum well, but at the cost of increased interface roughness leading to shorter scattering times~\cite{supplemental_hwang_limit_2008}. The mobility $\mu_\mathrm{e}$ and the the scattering time $\tau$ is determined by impurity content, which contributes to photon losses within the MQW structure and fundamentally limits the achievable signal boost in axion-induced fields, which has been experimentally demonstrated to reach $1700$\,ps~\cite{supplemental_chung_ultra-high-quality_2021} and may theoretically approach $4000$\,ps at the selected doping density~\cite{supplemental_chung_understanding_2022, supplemental_hwang_limit_2008}.
\subsection{High-field magnet}
\label{sub:magnet}
For axion detection, the signal $\Gamma_\text{signal} \propto \Bvec_\mathrm{ext}^2 A$ scales with the applied magnetic field strength $\Bvec_\mathrm{ext}$ and the area $A=W^2$ of the MQW structure. Like all axion haloscopes, a strong magnetic field is preferred, since it enhances the axion-induced signal. Additionally, a strong tunable magnetic field in the direction normal to the surface of the plasmonic cavity is preferred when targeting lower axion masses, closer to $\sim1$\,meV. Field homogeneity is especially critical for the normal component in the example configurations: the sharper the resonance at the ENZ point, the more sensitive the experiment becomes to spatial variations in $\boldsymbol{B}$. Figure~\ref{fig:boostvbfield} demonstrates how the boost factor for a fixed MQW and frequency peaks at the optimal magnetic field, while deviations due to inhomogeneity lead to a loss in coherence over the MQW and thus a reduction in sensitivity. For this reason, NMR magnets, such as the 36-T at the National High Magnetic Field Laboratory (NHMFL), are ideal, providing homogeneity at the 1-ppm level over centimeter-scale sample volumes, which ensures that the resonance condition is maintained across the entire device~\cite{supplemental_brandt_national_2001}. Configurations~1 and 2 utilize the 36-T magnet with an approximately 3-cm sample space, while Config.\,3 proposes an advanced 50-T magnet with a larger 5-cm sample space and 1-ppm  homogeneity, modeled after the 45-T NMR magnet also at NHMFL~\cite{supplemental_brandt_national_2001}; this may be achievable with reasonable technological advancements~\cite{supplemental_noauthor_opportunities_2005}. Lower homogeneity magnets may be used if the magnetic field component normal to the MQW surface is small, as the resonance is broader at lower fields. See ~\sref{sub:Binhom} for detailed simulations of the impact of inhomogeneity on detector response. 

Magnet and MQW parameters are co-optimized to ensure the resonance can be tuned across the target axion mass range by rotating the MQW within the fixed magnetic field, rather than ramping $\Bvec_\mathrm{ext}$ itself. 
Keeping the MQW structure fixed and varying the total magnetic field using a controlled current would not take full advantage of the total possible strength of the magnetic field, since scanning masses with resonances that occur at lower magnetic fields would lose out on maximizing the axion-induced electric field. 
The example \SQWARE{} will affix the MQW structure to a rotating probe within a constant magnetic field. This approach allows for continuous scan coverage and full utilization of the maximum available field. We note that the 36-T system at NHMFL is already equipped with a rotating probe. The time required to rotate the MQW structure (to sweep axion masses) is negligible compared to the total integration time per scan, as the detector remains operational during adjustments; the resonance frequency shifts smoothly with angle, enabling continuous coverage. This rotation-based scheme thus enables the experiment to exploit the full strength of the available magnetic field, maximize sensitivity, and efficiently scan the desired axion mass range.
\subsection{Focusing element}
\label{sub:lens_section}
Radiation emitted by the MQW structure will be guided onto the photodetector by a precisely aligned focusing assembly. Metalens have unique advantages with subwavelength thickness, high numerical apertures, and the ability to focus beyond the diffraction limit to improve coupling to photodetectors, reaching experimental focusing efficiencies up to around $70$\%~\cite{supplemental_legaria_highly_2021, supplemental_yang_high_2023}. Theoretical focusing efficiencies may reach close to 100\% with ideal diffractive or metasurface lenses or by using parabolic mirrors~\cite{supplemental_chen_3d-printed_2025, supplemental_lindlein_focusing_2019, supplemental_fleming_blazed_1997}; however, larger structures are required for high efficiencies, which may be a challenge if placing the lens and photodetector within the magnet with limited sample volume. This could be avoided with a coupler or by utilizing a magnet with windows. For the purposes of the example \SQWARE{} configurations' projected sensitivities, an ideal $100$\% focusing efficiency will be utilized for future projections.
For high-frequency (THz) operation, the MQW, lens, and photodetector are affixed to the rotating probe inside the magnet bore. 
If required by space or engineering constraints, a flexible optical fiber can guide photons from the rotating platform to a stationary photodetector situated outside the highest-field region, though this approach is generally feasible only at the upper end of the THz range where optical fibers exhibit acceptable transmission.
At lower frequencies, where suitable optical fibers are unavailable or excessively lossy, the MQW, lens, and photodetector must all be rigidly mounted within the magnet. 
Here, mechanical stability and precise optical alignment become paramount, as any misalignment could lead to a substantial loss in collection efficiency. 
Furthermore, in this configuration, the photodetector must maintain full sensitivity under high static magnetic fields and cryogenic temperatures, which is a nontrivial constraint for most superconducting technologies. We will see in the following section that quantum dot photodetectors are an ideal non-superconducting substitute and actually require high magnetic fields for tuning.
\subsection{Photodetector}
\label{sub:photodetector}
The photodetector is responsible for registering the rare THz photons produced via axion-photon conversion at the surface of the MQW structure. 
The performance of this photodetector is characterized primarily by its quantum efficiency $\eta$ and its dark count rate $\Gamma_\text{dark}$.
The photodetector must combine high quantum efficiency in the THz frequency range with an exceptionally low dark count rate, while remaining functional in high magnetic fields and cryogenic temperatures, and its signal must be digitized and recorded in real time for further analysis. Additionally, the photodetector must exhibit a bandwidth larger than the FWHM of the boost factor as a function of the resonance frequency to ensure continuous sweeping across all axion masses. 
Detecting single THz photons with high efficiency and low dark count rates remains an active area of research, often described as the ``THz technology gap.” 
Several promising approaches have emerged, each with advantages and trade-offs. 

Quantum dot photodetectors offer the lowest experimentally demonstrated dark count rates in the $1$-$10$\,meV ($0.25$-$2.5$\,THz) regime, with proven functionality in strong magnetic fields and cryogenic temperatures required for noise suppression~\cite{supplemental_komiyama_single-photon_2000,supplemental_astafiev_single-photon_2002,supplemental_komiyama_single-photon_2011,supplemental_kajihara_terahertz_2013,supplemental_shaikhaidarov_detection_2016}. These photodetectors also use high-mobility GaAs-AlGaAs 2DEG systems for minimal noise and include single quantum dot (SQD), double quantum dot (DQD), and double quantum well or charge-sensitive infrared detectors (CSIPs). 
In such devices, incident photons excite electron-hole pairs that generate charge signals read out via single-electron transistors or nearby 2DEGs. 
The spectral response tunable with a magnetic field, with Landau quantization, makes these devices highly specialized compared to conventional superconducting photodetectors. Dark counts (arising from charge fluctuations and intrinsic thermal activation~\cite{supplemental_astafiev_single-photon_2002})  down to $\Gamma_\text{dark} \sim 1$\,mHz have been achieved at quantum efficiencies $\eta \sim$ 1\% for photon energies of $2$-$6$\,meV and can theoretically reach $\mu$Hz. State-of-the-art bowtie antenna coupling has enabled quantum efficiencies up to 7\% in this regime, and with continued improvement in antenna design and device fabrication, efficiencies up to 20\% are achievable~\cite{supplemental_komiyama_single-photon_2011}. Recently, 35\% efficiency has been demonstrated for a mid-infrared CSIP using a germanium hemispherical mirror~\cite{supplemental_nakai_development_2024}.
For the \SQWARE{} example configurations, we take $\eta =$ 7, 20, and 35\% and $\Gamma_\text{dark} = 1, 1,$ and $0.1$\,mHz as realizable parameters, respectively.
\subsection{Optional mirror}
A key design choice in the \SQWARE{} is how to maximize the collection of axion-induced surface radiation emitted from the MQW structure, which inherently produces radiation from both of its parallel surfaces. 
One straightforward approach is to employ two independent optical paths, each with its own focuser and photodetector, on opposite sides of the MQW. 
While this method doubles the effective signal collection area, it also proportionally increases the total dark count rate, leading to only a modest improvement in sensitivity (by a factor of $2^{1/4} \approx 1.19$) due to the quartic scaling of sensitivity when background counts are included, as shown in Eq.~(6) in the main text.

An alternative and experimentally advantageous approach is to use a highly reflective conductor, or mirror, as a backing on one side of the MQW structure. 
As detailed in \sref{sub:radiation}, this configuration redirects the radiation that would have emerged from the rear surface back through the MQW, coherently adding it to the signal emitted from the front surface. 
In addition to concentrating the emitted power onto a single collection path (thereby eliminating the need for a second photodetector), the presence of a mirror modifies the electromagnetic boundary conditions and can further enhance the boost factor of the axion-induced electric field with thinner MQW structures. 
Analytically, this is encapsulated in \eref{eq:beta_mirror}, which demonstrates that a mirror-backed cavity can achieve a larger boost factor than the corresponding open geometry at optimal ENZ tuning. 
The mirror may be implemented as a metallic layer with a thickness at least comparable to the photon wavelength and a large refractive index (i.e., well within the perfect mirror limit). 
The mirror may be fabricated with standard substrate transfer processes on the surface opposite the primary radiative surface, or the MQW can be directly grown on a substrate with a high refractive index.

The use of a single photodetector and focuser simplifies the mechanical and optical alignment within the high-field magnet environment and reduces space requirements, which is critical for a large-area MQW and compact cryogenic setups.
The enhancement of the boost factor with the addition of the mirror allows for optimization of the MQW thickness and composition, potentially requiring fewer quantum wells to achieve a desired sensitivity, thus shortening growth and fabrication time.
The conductor must not introduce significant loss at the target THz frequencies and must remain compatible with the overall cryogenic and high-magnetic-field environment. 
Standard choices include gold, silver, or aluminum films, whose reflectivities approach unity in the relevant frequency range. 
This design choice can be revisited as further simulations and prototypes clarify the comparative gains of the mirror-backed architecture.

\subsection{Calibration and signal acquisition}
\label{calibration}
\begin{figure}[H]
    \centering
\includegraphics[height=0.30\linewidth]{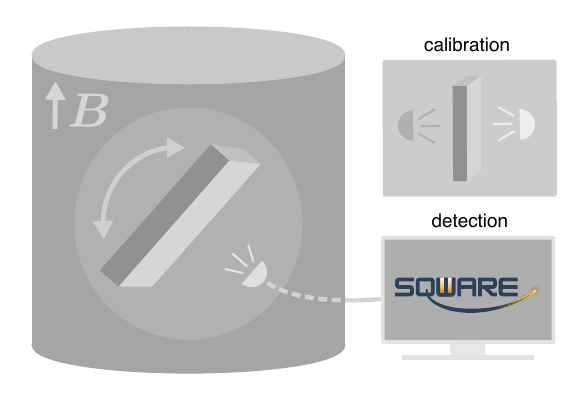}\\
    \caption{Data Acquisition and Calibration: The MQW may be rotated within a strong magnetic field with an affixed photodetector to collect the surface radiation. Alternatively, a cavity or optical fiber could be used to direct light to an external photodetector. The signal is read out through a dedicated computer. The calibration of the material resonance may be done prior to or \textit{in~situ} during the experiment, by measuring the reflection or transmission by an external THz source.}
    \label{daq}
\end{figure}

The successful operation of \SQWARE{} depends not only on maximizing axion-induced signal power, but also on precise calibration and robust signal acquisition protocols that reliably distinguish potential axion signals from backgrounds. 
The design of the acquisition system varies somewhat depending on whether the experiment is optimized for low or high THz frequencies, as well as for the physical configuration of the MQW structure, focusing element, and photodetector within the magnet. 
Signal photon events are recorded in real time, with individual photon arrival times and pulse heights stored for offline analysis. 
Background rates are established \textit{in~situ} by recording data with the magnetic field turned off or with the MQW rotated out of resonance (i.e., at an angle where the resonance frequency is not within the bandwidth of the photodetector). 
These ``dark runs” enable robust background subtraction to account for any drifts in the baseline dark count rate and the identification of non-axionic backgrounds, including ambient photon events.
The overall data acquisition (DAQ) system is controlled by a dedicated computer, interfaced to the photodetector and the magnetic field control/rotation system.
(see Figure~\ref{daq}). 
Synchronized control allows for automated scanning over field orientation and real-time logging of system status and environmental parameters (magnetic field, temperature, position, etc.), ensuring traceability and reproducibility of all scans. 

Accurate calibration is essential for both locating the ENZ resonance and establishing the frequency and angular response of the MQW structure under operational conditions. 
For this purpose, an external tunable THz source can be coupled into the system to illuminate the MQW structure, and the transmission or reflection ellipsometry can be performed to directly yield the effective permittivity as a function of frequency and magnetic field. Using the experimentally measured permittivity along with the material structure and magnetic field profile in simulation will accurately determine the boost as a function of frequency and magnetic field and thus the resonance quality factor of the MQW cavity.
Flexibility in optical and mechanical design, together with \textit{in~situ} calibration and advanced noise reduction, will enable the unambiguous identification (or exclusion) of axion-induced signals in the presence of realistic backgrounds.


   
%

\end{document}